\documentclass[nofootinbib, twocolumn,superscriptaddress,preprintnumbers,letterpaper,nofootinbib,natbib]{revtex4-1}

\usepackage[english]{babel}

\usepackage{graphicx}
\usepackage{dcolumn}
\usepackage{bm}
\usepackage[breaklinks,colorlinks,urlcolor=blue,citecolor=blue,linkcolor=magenta]{hyperref}
\usepackage{epsfig}
\usepackage[usenames,dvipsnames]{xcolor}
\usepackage{multirow}
\usepackage{capt-of}
\usepackage{siunitx}
\usepackage{graphicx}
\usepackage{slashed}
\usepackage{tabularx}
\usepackage{multirow}
\usepackage{braket}
\usepackage{amsfonts}
\usepackage{mathtools}
\usepackage{mathrsfs}
\usepackage{bbold}
\usepackage[compat=1.1.0]{tikz-feynman}
\usepackage{subfiles}
\usepackage{comment}

\usepackage{amsmath}
\usepackage{cleveref}

\usepackage{float}

\usepackage{slashed}
\usetikzlibrary{positioning}
\usepackage[compatibility=false]{caption}
\usepackage{subcaption}

\usepackage[normalem]{ulem}
\usepackage[shortlabels]{enumitem}  

\usepackage{fontawesome}
\setcitestyle{sort&compress}

\usepackage{float}

\newcommand{\coeff}[2]{ \mathcal{C}_{#1} ^{#2} }

\newcommand{\order}[1] {\mathcal{O }\left( #1 \right)}
\newcommand{\yuk}[1]{{Y}_{#1}}

\newcommand{\hc}{\text{H.c.}}

\pgfmathsetmacro\sizedot{1.1}
\pgfmathsetmacro\sizesqdot{1.5}
\pgfmathsetmacro\sizecrodot{1.0}

{\left\lbrace\begin{array}{@{}l@{}}}%
{\end{array}\right.}
\usepackage{slashed}

\begin{document}
\preprint{P3H-23-080} \preprint{KA-TP-22-2023} \preprint{TTP23-054} 

\title{On $\gamma_5$ schemes and the interplay of SMEFT operators in the Higgs-gluon coupling
}
\author{Stefano Di Noi}
\email{stefano.dinoi@phd.unipd.it} 
\author{Ramona Gr\"ober}%
 \email{ramona.groeber@pd.infn.it}
 \affiliation{%
 Dipartimento di Fisica e Astronomia ``G. Galilei", Universit\`a di Padova, Italy, \\ and Istituto
Nazionale di Fisica Nucleare, Sezione di Padova, I-35131 Padova, Italy
}
\author{Gudrun Heinrich}
\email{gudrun.heinrich@kit.edu}
\author{Jannis Lang}
\email{jannis.lang@kit.edu}
 \affiliation{%
 Institute for Theoretical Physics, Karlsruhe Institute of Technology (KIT), D-76131 Karlsruhe, Germany
}\author{Marco Vitti}
  \email{marco.vitti@unipd.it}
  \affiliation{%
 Dipartimento di Fisica e Astronomia ``G. Galilei", Universit\`a di Padova, Italy, \\ and Istituto
Nazionale di Fisica Nucleare, Sezione di Padova, I-35131 Padova, Italy
}
 \affiliation{%
 Institute for Theoretical Particle Physics, Karlsruhe Institute of Technology (KIT), D-76131 Karlsruhe, Germany
}
\affiliation{%
 Institute for Astroparticle Physics, Karlsruhe Institute of Technology (KIT), D-76344 Eggenstein-Leopoldshafen, Germany
}
\date{\today}

\begin{abstract}
We calculate the four-top quark operator contributions to Higgs production via gluon fusion in the Standard Model Effective Field Theory. The four-top operators enter for the first time via two-loop diagrams. Due to their chiral structure they contain $\gamma_5$, so special care needs to be taken when using dimensional regularisation for the loop integrals. We use two different schemes for the continuation of $\gamma_5$ to $D$ space-time dimensions in our calculations and present a mapping for the parameters in the two schemes. This generically leads to an interplay of different operators, such as four-top operators, chromomagnetic operators or Yukawa-type operators at the loop level. We validate our results by examples of matching onto UV models.
\end{abstract}

\maketitle

\section{\label{sec:intro}Introduction}
With the increasing precision in the measurement of the Higgs boson couplings, the Higgs sector has become a probe of physics beyond the Standard Model (SM). In the absence of a clear signal of new physics, potential deviations from the SM can be described as model-independently as possible by means of an effective field theory (EFT). Under the assumption that the Higgs field transforms as an $SU(2)_L$ doublet as in the SM, heavy new physics can be described by the SM effective field theory (SMEFT), see Refs.~\cite{BUCHMULLER1986621, dim6smeft,Brivio:2017vri,Isidori:2023pyp}. In this theory, new physics effects are described by higher-dimensional operators suppressed by some large mass scale $\Lambda$. 

In this paper we consider a subset of the possible dimension-six operators, namely the four-top quark operators, and comment on their connection to other SMEFT operators. Four-top operators are generically difficult to probe experimentally, as direct probes require the production of four top quarks. Limited by the large phase space required, four-top quark production remains a rather rare process, with a SM cross section of only about 12 fb including next-to-leading (NLO) QCD and NLO electroweak (EW) corrections for $\sqrt{s}=13$ TeV \cite{Bevilacqua:2012em,Frederix:2017wme,Jezo:2021smh}. Limits have been mainly presented for $\mathcal{O}(1/\Lambda^4)$ contributions in the matrix element squared~\cite{Hartland:2019bjb}, but can potentially also be derived from the $\mathcal{O}(1/\Lambda^2)$ interference with the SM only~\cite{Aoude:2022deh,ATLAS:2023ajo}. In particular, four-top production has been recently observed by ATLAS and CMS \cite{CMS:2023ftu,ATLAS:2023ajo}, with bounds ranging from $\sim 1$ to $\sim 7$ $\text{TeV}^{-2}$ on the absolute values of the four-top Wilson coefficients.

On the other hand, complementary bounds on the four-top operators can be obtained indirectly, hence by considering loop effects on other observables. 
The four-top operator contributions (via one-loop corrections) to $t\bar{t}$ production were discussed in Ref.~\cite{Degrande:2020evl} with the conclusion that the effects on the total cross section are small due to cancellations between different phase-space regions and due to suppressed interference with the SM QCD amplitude.  A differential analysis has not been performed yet.

Furthermore, Ref.~\cite{Alasfar:2022zyr} showed that loop contributions from four-top operators in Higgs production processes can be important, not only as probes of the relative Wilson coefficients but also because in the presence of four-top operators possible limits on the trilinear Higgs self-coupling derived from electroweak corrections to single Higgs production \cite{McCullough:2013rea, Gorbahn:2016uoy, Degrassi:2016wml, Bizon:2016wgr, Maltoni:2017ims, DiVita:2017eyz, Degrassi:2019yix} can become less restrictive.
First efforts to constrain the trilinear Higgs self-coupling via single Higgs production have already been performed by the experimental collaborations \cite{ATLAS:2022jtk,CMS:2023gjz}. 

We are going to reconsider the $gg\to h$ computation from Ref.~\cite{Alasfar:2022zyr}, which included effects from four-top operators within the SMEFT, using two different schemes for the continuation of $\gamma_5$ to $D=4-2\epsilon$ space-time dimensions. 
While the leading poles of loop integrals are scheme-independent, cancellations of these poles with scheme-dependent $\order{\epsilon}$ terms, resulting from the Dirac algebra in dimensional regularisation, will lead to scheme-dependent finite parts. It should be stressed that, in this context, the finite terms can be of the same order as the logarithmically enhanced ones (as shown in Ref.~\cite{Alasfar:2022zyr}), thus they are phenomenologically relevant. 
Since four-top operators contribute to $gg\to h$ via two-loop diagrams, the finite terms are expected to be scheme-dependent. Moreover, we find a divergence which depends on the scheme, signaling a scheme-dependent anomalous dimension. We describe in detail how such divergence can be traced back to a finite term (that is expected to be scheme-dependent) in one of the one-loop subamplitudes entering the computation.
We also review the results in na\"ive dimensional regularisation (NDR) \cite{CHANOWITZ1979225} with respect to the ones obtained in Ref.~\cite{Alasfar:2022zyr} and we discuss various subtleties that arise in the comparison with the \textit{Breitenlohner-Maison-'t Hooft-Veltman} scheme (BMHV) \cite{THOOFT1972189,Breitenlohner:1977hr} for the treatment of $\gamma_5$. We refer the reader to  Ref.~\cite{Boughezal:2019xpp} for another comparison of different $\gamma_5$ schemes within the SMEFT.

Furthermore, we point out that building the SMEFT expansion on the counting of the canonical dimension alone can lead to inconsistencies, as has been explained in Ref.~\cite{Buchalla:2022vjp}. In a counting scheme that in addition takes into acccount whether an operator is potentially loop-generated, the four-top operators and the chromomagnetic operator enter the Higgs-gluon coupling at the same order \cite{Arzt:1994gp,Giudice:2007fh,Contino:2013kra,Buchalla:2014eca,Englert:2019rga,Muller:2021jqs,Buchalla:2022igv,Guedes:2023azv} and therefore should not be considered in isolation. 

Our paper is structured as follows: in Sec.~\ref{sec:operators} we introduce the operators considered in our analysis and we fix our notation. In Sec.~\ref{sec:gamma5} we discuss different schemes for the $D$-dimensional continuation of $\gamma_5$. 
Section~\ref{sec:preliminary} is devoted to the computation of one-loop subamplitudes required to obtain the result for the $gg\to h$ amplitude including the operators given in Sec.~\ref{sec:operators}.
The two different schemes are then used for the computation of the $gg\to h$ rate presented in Sec.~\ref{sec:calculation}. We also discuss how the scheme-dependence of the parameters of the theory compensates for the scheme-dependence of the matrix elements, providing a scheme-independent physical result. In Sec.~\ref{sec:matching} we validate our approach by means of a matching with two simple models. In Sec.~\ref{sec:moreinterplay} we briefly show that a non trivial interplay exists not only in the case of four-top operators, as detailed in this work, but also when other operators containing chiral vertices are involved.
In App.~\ref{app:htobb} we show the result we obtain for $\Gamma \left( h \to \bar{b}b \right)$ as a side-product of our analysis, commenting also in this case about the scheme-independence of the result. In App.~\ref{app:RGE} we discuss the relation between the counterterms and the anomalous dimension matrix, highlighting some subleties that arise when dimensional regularisation is used. In App.~\ref{app:results} we report the scheme-independent part of the $gg\to h$ amplitude and in App.~\ref{app:FR} we give the Feynman rules needed for our computation.
\section{\label{sec:operators} Setup}
If the new physics scale $\Lambda$ is assumed to be much larger than the electroweak scale, new physics can be described in terms of an EFT. In this paper we use the SMEFT, where all SM fields transform under the SM symmetries, including the scalar field $\phi$ which contains the Higgs boson. 
At dimension-five level there is only the lepton-number violating ``Weinberg'' operator responsible for Majorana mass generation of neutrinos \cite{PhysRevLett.43.1566}, so the dominant new physics effects relevant in collider physics are described by dimension-six operators:
\begin{equation}
\mathcal{L}_{\mathrm{\mathcal{D}=6}}=\mathcal{L}_{\mathrm{SM}}+\frac{1}{\Lambda^2}\sum_{i}\coeff{i}{}\mathcal{O}_i\,,
\end{equation}
where $\mathcal{O}_i$ denotes every possible non-redundant combination of SM fields with mass dimension six that preserves the symmetries of the SM. 
A complete basis of dimension-six operators was presented for the first time in Ref.~\cite{dim6smeft}, the so-called \textit{Warsaw basis}, that we will adopt in the following. In the Warsaw basis redundant operators are eliminated making use of field redefinitions, integration-by-part identities and Fierz identities.

We are mostly interested in the effect of the four-top operators on Higgs production via gluon fusion (as well as the Higgs decay to gluons). The operators that lead to four-top interactions are given by
\begin{equation} \label{eq:Lag4t}
\begin{split}
\mathcal{L}_{\text{4t}} &= \frac{\coeff{QQ}{(1)} }{\Lambda^2}\left(\bar{Q}_L \gamma_\mu Q_L\right)  \left(\bar{Q}_L \gamma^\mu Q_L \right) \\
&+\frac{\coeff{QQ}{(3)}}{\Lambda^2} \left(\bar{Q}_L \tau^I \gamma_\mu Q_L \right) \left(\bar{Q}_L \tau^I \gamma^\mu Q_L \right)  \\
&+ \frac{\coeff{Qt}{(1)}}{\Lambda^2}  \left(\bar{Q}_L \gamma_\mu Q_L \right) \left(\bar{t}_R \gamma^\mu t_R \right) \\
&+\frac{\coeff{Qt}{(8)}}{\Lambda^2} \left(\bar{Q}_L T^A\gamma_\mu Q_L \right) \left(\bar{t}_R T^A \gamma^\mu t_R \right) \\
&+ \frac{\coeff{tt}{}}{\Lambda^2} \left(\bar{t}_R  \gamma_\mu t_R \right) \left(\bar{t}_R \gamma^\mu t_R \right)\,.
\end{split}
\end{equation}
The field $Q_L$ stands here for the $SU(2)_{L}$ doublet of the third quark generation, $t_R$ for the right-handed top quark field. The $SU(3)_C$ generators are denoted as $T^A$ while $\tau^I$ are the Pauli matrices. We assume all the Wilson coefficients to be real, since we are not interested in CP-violating effects.

The operators in Eq.~\eqref{eq:Lag4t} contribute to the $gg \to h$ amplitude via two-loop diagrams. At one-loop and tree-level, respectively, the following operators contribute to the (CP-even) Higgs-gluon coupling
\begin{equation}
\begin{split}
\mathcal{L}_{\text{2t}}= &\left[\frac{\coeff{t\phi}{}}{\Lambda^2} (\bar{Q}_L \tilde{\phi} t_R )\phi^{\dagger}\phi  +\frac{\coeff{tG}{}}{\Lambda^2} \bar{Q}_L \sigma^{\mu\nu} T^A t_R \tilde{\phi}G_{\mu\nu}^{A}+ \hc \right], \\
\mathcal{L}_{\text{s}}= &\frac{\coeff{\phi G}{}}{\Lambda^2} \phi^{\dagger}\phi G_{\mu \nu}^A G^{A\mu \nu} ,
\end{split}\label{eq:Lag2ts}
\end{equation}
where $G_{\mu\nu}^A=\partial_\mu G_\nu^A-\partial_\nu G_\mu^A-g_s f^{ABC}G_\mu^B G_\nu^C$ is the gluon field strength tensor, $\tilde{\phi}=i\tau^2 \phi^*$ and $\sigma_{\mu\nu}=i/2[\gamma_{\mu},\gamma_{\nu}]$. The operator in the second term of $\mathcal{L}_{\mathrm{2t}}$ is known as the chromomagnetic operator and will have a central role in this paper, as detailed in the following.
\par
To summarise our EFT setup, our Lagrangian reads:

\begin{equation} \label{eq:totLag}
\mathcal{L}_{\mathrm{\mathcal{D}=6}}=\mathcal{L}_{\mathrm{SM}}+\mathcal{L}_{\text{4t}}+\mathcal{L}_{\text{2t}}+\mathcal{L}_{\text{s}}.
\end{equation}
We follow Ref.~\cite{Jenkins:2017jig} for what concerns the conventions in $\mathcal{L}_{\mathrm{SM}}$,
\begin{equation}\label{eq:SMlag}
\begin{aligned}
			\mathcal{L}_{\textrm{SM}} =&  -\frac{1}{4} G_{\mu \nu}^A G^{A\mu \nu}
			-\frac{1}{4} W_{\mu \nu}^I W^{I\mu \nu}
			-\frac{1}{4} B_{\mu \nu}B^{\mu \nu} \\
			&+ \sum_{\psi} \bar{\psi} i \slashed{D} \psi 
			+ (D_\mu \phi)^\dagger (D^\mu \phi) \\ &- \lambda \left(\phi^\dagger \phi - \frac{1}{2}v^2 \right)^2 -
 \yuk{u} \tilde{\phi}^\dagger \bar{u}_R  Q_L + \hc .
\end{aligned}
\end{equation}
The gauge covariant derivative is $D_{\mu} = \partial_{\mu} + i g' \mathsf{y}  B_\mu+ i g \tau^I W_{\mu}^I + i g_s T^A G_{\mu}^A $,  $\mathsf{y}$ being the hypercharge.
When spontaneous symmetry breaking occurs ($\phi=(1/\sqrt{2})(
    0 ,(v+h))^T$ in the unitary gauge) one has:
\begin{equation}
\mathcal{L}_{\mathcal{D}=6} \supset - m_t \bar{t} t - g_{h \bar{t}t} h \bar{t} t,
\end{equation}
where the top mass and the $h\bar{t}t$ coupling are modified according to 
\begin{equation}
\begin{aligned} \label{eq:ghttmtBF}
m_t &= \frac{v}{\sqrt{2}} \left( \yuk{t} -\frac{v^2}{2} \frac{\coeff{t \phi}{}}{\Lambda^2} \right),  \\ 
g_{h \bar{t}t} &= \frac{1}{\sqrt{2}} \left( \yuk{t} -\frac{3 v^2}{2} \frac{\coeff{t \phi}{}}{\Lambda^2} \right)=\frac{m_t}{v} - \frac{v^2}{\sqrt{2}} \frac{\coeff{t \phi}{}}{\Lambda^2}.
\end{aligned}
\end{equation}
This establishes a connection between $m_t,g_{h \bar{t}t}$ (broken phase) and $\yuk{t},\coeff{t \phi}{}/\Lambda^2$ (unbroken phase).

\section{\label{sec:gamma5} Continuation schemes for $\gamma_5$ to $D$ dimensions}

To deal with loop integration, we have to choose a regularisation scheme. We employ dimensional regularisation, see Refs.~\cite{Gnendiger:2017pys,Belusca-Maito:2023wah} for a review. 
Due to the presence of four-fermion operators with different chiralities, $\gamma_5$ matrices will be present in our loop computations. As well known, the treatment of $\gamma_5$ in dimensional regularisation is highly non-trivial, as $\gamma_5$ is an intrinsically four-dimensional object (e.g.~\cite{Jegerlehner:2000dz}).
In this paper, we will consider two different schemes for the $\gamma_5$ matrix in dimensional regularisation with $D=4-2\epsilon$: na\"ive dimensional regularisation and the Breitenlohner-Maison-t'Hooft-Veltman scheme. 

\subsection{Na\"ive Dimensional Regularisation}
The NDR scheme assumes that the usual anti-commutation relations valid in four dimensions hold also in $D$ dimensions
\begin{equation}
\{\gamma_{\mu}, \gamma_{\nu} \}=2 g_{\mu\nu}\,, \hspace*{0.5cm} \{\gamma_{\mu}, \gamma_5 \}=0\,, \hspace*{0.5cm}\gamma_5^2=\mathbb{1}\,.
\end{equation}
This is inconsistent with the cyclicity of the trace. Assuming that the usual four-dimensional relation
\begin{equation}
\text{Tr}[\gamma_{\mu}\gamma_{\nu}\gamma_{\rho}\gamma_{\sigma}\gamma_5]=-4i\epsilon_{\mu\nu\rho\sigma}
\end{equation}
holds, leads to 
\begin{equation}
\text{Tr}[\gamma_{\mu_1} \gamma_{\mu_2}..\gamma_{\mu_{2n}}\gamma_5]=\text{Tr}[\gamma_{\mu_2}..\gamma_{\mu_{2n}}\gamma_5\gamma_{\mu_1}]+\mathcal{O}(\epsilon), 
\end{equation}
for $n\ge 3$. 
The cyclicity is hence no longer preserved and the computation of a Feynman diagram depends on the starting point of reading in a fermion trace. As was shown in Refs.~\cite{Korner:1991sx, Kreimer:1993bh}, the NDR scheme in presence of Dirac traces with an odd number of $\gamma_5$ matrices and at least six $\gamma$-matrices only leads to consistent results if the reading point is fixed univocally for all Feynman diagrams.\footnote{It was shown recently in Ref.~\cite{Chen:2023ulo} that in a computation of the singlet axial-current operator at $\mathcal{O}(\alpha_s^3)$ between two gluons and the vacuum a revised version of the scheme of Refs.~\cite{Korner:1991sx, Kreimer:1993bh} becomes necessary.} 

\subsection{Breitenlohner-Maison-'t Hooft-Veltman Scheme}
The BMHV scheme divides the algebra in a four-dimensional part and a $(D-4)$-dimensional one by defining
\begin{align}
\gamma_\mu^{(D)}&=\gamma_\mu^{(4)}+\gamma_\mu^{(D-4)},\\
\{\gamma_{\mu}^{(4)}, \gamma_5 \}&=0, \quad
[\gamma_{\mu}^{(D-4)}, \gamma_5 ]=0.
\end{align}

For the vertices involving chiral projectors we use the following rule, valid in the BMHV scheme:
\begin{equation}
    \gamma_\mu^{(4)}(1 \mp \gamma^5) \to \frac{1}{2} (1 \pm \gamma^5) \gamma_\mu^{(D)}(1 \mp \gamma^5),
    \label{eq:symmrule}
\end{equation}
which is the most symmetric choice and preserves chirality of the external fields in $D$ dimensions (see e.g.~Refs.~\cite{Ciuchini:1993vr, Belusca-Maito:2023wah, Cornella:2022hkc}).

The BMHV continuation scheme breaks explicitly the chiral simmetry. For this reason, symmetry-restoring finite counterterms may be required, as described in Ref.~\cite{Larin:1993tq} for QCD corrections and in Ref.~\cite{Boughezal:2019xpp} in the case of the SMEFT. For the purpose of this paper, such counterterms are not required. We have verified that the Lorentz structure of our final result in both schemes is the one expected from gauge invariance ($L^{\mu_1 \mu_2}$, see Eq.~\eqref{eq:Lorentz}), which we consider a consistency check of the result in the BMHV scheme.

\section{Scheme-dependent finite mixing at one-loop order} \label{sec:preliminary}
In this section we comment on the interplay between the four-top operators and other operators entering Eq.~\eqref{eq:totLag}. This interplay will be important in the discussion of single Higgs production in the next section. 

In particular, we want to highlight two points. The first one is that there is a finite mixing between the four-top and other operators, coming already from one-loop diagrams, as shown below. This fact implies that it would be inconsistent to study the contribution coming from four-top operators in isolation. The second point is that the above mixing, being finite, depends on the $\gamma_5$ scheme employed. When combining the one-loop subamplitudes in two-loop diagrams, in principle this could lead to divergent terms that are scheme-dependent. However, provided that both schemes are used consistently,
the physical result for the complete two-loop amplitude is expected to be scheme-independent.

\begin{figure}[t!]
    \begin{minipage}{0.49\linewidth}

         \begin{tikzpicture}[baseline=(4F)]
            \begin{feynman}[small]
                \vertex  (g1)  {$g$};
                \vertex (gtt1) [dot, scale=0.3, right=25 pt of g1] {};
                \vertex  (4F) [square dot, scale=0.01, right=25pt of gtt1, color=white] {};
                \vertex (c1) [crossed dot, scale=0.6, left=2.5pt of 4F] {};
                \vertex (c2) [crossed dot, scale=0.6, right=2.5pt of 4F] {};
                \vertex  (t1) [above right=25pt of 4F] {$t$};
                \vertex (t2) [below right=25pt of 4F] {$t$};

                \diagram* {
                    (g1)  -- [gluon] (gtt1),  
                    (gtt1) -- [anti fermion, half right] (c1) --[anti fermion, half right] (gtt1),                
                    (t1) -- [anti fermion] (c2) -- [anti fermion] (t2), 
                };
            \end{feynman}
        \end{tikzpicture}
       \subcaption{}
        \label{fig:schannel}
    \end{minipage}%
    \hfill
    \begin{minipage}{0.49\linewidth}
        \centering
       \begin{tikzpicture}[baseline=(4F)]
            \begin{feynman}[small]
                \vertex  (g1)  {$g$};
                \vertex (gtt1) [dot, scale=0.3, right=25 pt of g1] {};
                \vertex  (4F) [square dot, scale=0.01, right=20pt of gtt1, color=white] {};
                \vertex (c1) [crossed dot, scale=0.6, above=2.5pt of 4F] {};
                \vertex (c2) [crossed dot, scale=0.6, below=2.5pt of 4F] {};
                \vertex  (t1) [above right=25pt of 4F] {$t$};
                \vertex (t2) [below right=25pt of 4F] {$t$};

                \diagram* {
                    (g1)  -- [gluon] (gtt1),  
                    (gtt1) -- [anti fermion, half right] (c2) --[anti fermion] (t2), 
                    (gtt1) -- [fermion, half left] (c1) --[fermion] (t1),
                };
            \end{feynman}
        \end{tikzpicture}
        \subcaption{}
        \label{fig:tchannel}
    \end{minipage}
    \raggedleft
    \caption{The two possible contractions within four-fermion operators where all the fermions are equal: (a) closed fermion line yielding a trace; (b) open fermion line without any traces.}
    \label{fig:contractions}
\end{figure}
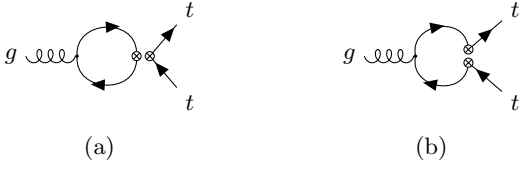

Direct evaluation of the contribution of the four-top operators to the $g \to \bar{t}t$ amplitude gives a contribution proportional to an insertion of the chromomagnetic operator. 
Pictorially, this can be represented as follows
\begin{equation}
        \begin{tikzpicture}[baseline=(4F)]
            \begin{feynman}[small]
                \vertex  (g1)  {$g$};
                 \vertex (gtt1) [dot, scale=\sizedot, right= 30 pt of g1] {};
                \vertex  (4F) [square dot,scale=\sizesqdot,right = 20 pt of gtt1, color = red] {};
                \vertex  (t1) [above right= 25 pt of 4F] {$t$};
                \vertex (t2) [below right= 25 pt of 4F] {$t$};

                \diagram* {
                    (g1)  -- [gluon] (gtt1),  
                    (gtt1) -- [anti fermion, half right] (4F) 
                    -- [anti fermion,half right] (gtt1),
                    (t2) -- [fermion]  (4F) -- [fermion] (t1)
                };
            \end{feynman}
        \end{tikzpicture} = \frac{\coeff{Qt}{(1)}-\frac{1}{6} \coeff{Qt}{(8)}}{\coeff{tG}{}} K_{tG}  \times 
                \begin{tikzpicture}[baseline=(4F)]
            \begin{feynman}[small]
                \vertex  (g1)  {$g$};
                 \vertex (gtt1) [square dot, scale=\sizesqdot, right= 30 pt of g1, color=blue] {};
                \vertex  (t1) [above right= 25 pt of gtt1] {$t$};
                \vertex (t2) [below right= 25 pt of gtt1] {$t$};

                \diagram* {
                    (g1)  -- [gluon] (gtt1),  
                    (t2) -- [fermion] (gtt1) -- [fermion] (t1),};
            \end{feynman}
        \end{tikzpicture}
        \label{eq:diag_tg},
\end{equation}
where the red and blue square dots denote an insertion of four-top and chromomagnetic operators, respectively. \newline 
The value of $K_{tG}$ in Eq.~\eqref{eq:diag_tg} depends on the $\gamma_5$ scheme. We find
\begin{equation}
K_{tG} =
\begin{cases}
\frac{\sqrt{2} m_t g_s}{16 \pi^2
 v} & \text{(NDR)}\\
0 & \text{(BMHV)}.
\end{cases}
\label{eq:Ktg}
\end{equation}
We note that Eq.~\eqref{eq:diag_tg} holds only when the gluon is on shell. In this case, only one of the two possible contractions of the fermion lines, namely the one in Fig.~\ref{fig:tchannel} featuring an open fermion line, gives a non-vanishing contribution. Therefore, the difference between the two schemes in Eq.~\eqref{eq:Ktg} does not arise from a trace in Dirac space and cannot be related to trace ambiguities \cite{Korner:1991sx}.

When we consider other one-loop amplitudes with four-top operator insertions, which will enter as subamplitudes in the $gg \to h$ computation, we find again that the finite contributions are scheme-dependent, whereas the divergent parts are equal in the two schemes.  
In particular, the diagrammatic relation concerning the finite part of the four-top contribution to the Higgs-top coupling is

\begin{equation} 
\begin{aligned}
        \begin{tikzpicture}[baseline=(4F)]
            \begin{feynman}[small]
                \vertex  (g1)  {$h$};
                 \vertex (gtt1) [dot, scale=\sizedot, right= 25 pt of g1] {};
                \vertex  (4F) [square dot,scale=\sizesqdot,right = 20 pt of gtt1, color = red] {};
                \vertex  (t1) [above right= 25 pt of 4F] {$t$};
                \vertex (t2) [below right= 25 pt of 4F] {$t$};

                \diagram* {
                    (g1)  -- [scalar] (gtt1),  
                    (gtt1) -- [anti fermion, half right] (4F) 
                    -- [anti fermion,half right] (gtt1),
                    (t2) -- [fermion]  (4F) -- [fermion] (t1)
                };
            \end{feynman}
        \end{tikzpicture} \Biggl |_\text{FIN} &= \frac{1}{\Lambda^2} \, \left( \coeff{Qt}{(1)} + \frac{4}{3} \coeff{Qt}{(8)}\right) \\ &\times (B_{{h\bar{t}t}}+K_{{h\bar{t}t}}) \times 
                \begin{tikzpicture}[baseline=(4F)]
            \begin{feynman}[small]
                \vertex  (g1)  {$h$};
                 \vertex (gtt1) [dot, scale=\sizedot, right= 25 pt of g1, color=black] {};
                \vertex  (t1) [above right= 25 pt of gtt1] {$t$};
                \vertex (t2) [below right= 25 pt of gtt1] {$t$};

                \diagram* {
                    (g1)  -- [scalar] (gtt1),  
                    (t2) -- [fermion] (gtt1) -- [fermion] (t1),};
            \end{feynman}
        \end{tikzpicture},
        \label{eq:diag_ghtt}
        \end{aligned}
\end{equation}
where we find
\begin{equation}
K_{h\bar{t}t} =
\begin{cases}
\frac{(m_h^2-6m_t^2)}{16 \pi^2} & \text{(NDR)}\\
0 & \text{(BMHV)}.
\end{cases}
\label{eq:Khtt}
\end{equation}
$B_{h\bar{t}t}$ is scheme-independent and can be expressed as
\begin{equation}
\begin{aligned}
B_{h\bar{t}t} &= \frac{m_t^2}{4 \pi ^2 \tau } \times \Bigg(  -2 \beta ^3 \log \left(\frac{\beta -1}{\beta +1} \right)\\
&  +(3 \tau -2) \log \left(\frac{\tilde{\mu}^2 }{m_t^2}\right)+5 \tau -4 \Bigg),
\end{aligned}
\end{equation}
with 
\begin{equation}
 \qquad \beta= \sqrt{1- \tau} , \qquad \tau =\frac{4 m_t^2}{m_h^2}
\label{eq:beta}
\end{equation}
and with $\tilde{\mu}^2 = 4 \pi \mu^2 e^{-\gamma_{E}}  $.
We note that $B_{h \bar{t}t}$ and the analogous $B$ terms in this paper are scheme-independent once a convention to identify $K_{h \bar{t}t}$ is defined. For example, in this section we choose the $B$ terms such that the $K$-terms vanish in BMHV. However, this definition is totally arbitrary and does not affect the final results. What is relevant for our purpose is the difference between $K$-terms in different schemes, which is insensitive to the convention chosen.

\begin{figure}[h]
\centering
    \begin{subfigure}[t]{0.45\linewidth}
\centering
\begin{tikzpicture} 
            \begin{feynman}[small]
                \vertex  (g1)  {$g$};
                \vertex  (gtt1) [square dot, scale=\sizesqdot,right= of g1,color=blue] {};
                \vertex (htt) [dot,scale=\sizedot,below right = of gtt1]  {};
                \vertex  (gtt2) [dot,scale=\sizedot,below left = of htt] {};
                \vertex  (g2) [left=of gtt2]  {$g$};
                \vertex (h) [right = of htt] {$h$};

                \diagram* {
                    (g1)  -- [gluon] (gtt1),
                    (g2) -- [gluon] (gtt2),
                    (h)  -- [scalar] (htt),
                    (gtt1) -- [fermion] (htt) 
                    -- [fermion] (gtt2)
                     -- [fermion] (gtt1), 
                    
                };
            \end{feynman}
        \end{tikzpicture}
        \caption{Triangle topology.}
        \end{subfigure}
\begin{subfigure}[t]{0.45\linewidth}
\centering
\begin{tikzpicture} 
            \begin{feynman}[small]
                \vertex  (g1)  {$g$};
                \vertex  (gtt1) [square dot, scale=\sizesqdot,right= of g1,color=blue] {};
                 \vertex (htt) [dot,scale=0.01,below right = of gtt1,color=white]  {};
                \vertex  (gtt2) 
                [dot,scale=\sizedot,below left = of htt] {};
                \vertex  (g2) [left=of gtt2]  {$g$};
                \vertex (h) [right = of gtt1] {$h$};

                \diagram* {
                    (g1)  -- [gluon] (gtt1),
                    (g2) -- [gluon] (gtt2),
                    (h)  -- [scalar] (gtt1),
                    (gtt1) -- [fermion,quarter right] (gtt2)
                     -- [fermion, quarter right] (gtt1), 
                    
                };
            \end{feynman}
        \end{tikzpicture}
        \caption{Bubble topology.}
        \end{subfigure}

        \caption{Contribution to the Higgs-top quark coupling with a single insertion of the chromomagnetic operator (blue square dot).}\label{fig:cm_1loop}
    \end{figure}
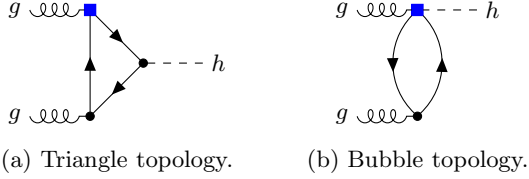

Regarding the corrections to the top quark propagator we find that only the mass term gets corrected, see App.~\ref{app:FR}. Diagrammatically, we have

\begin{equation}\begin{aligned}
\begin{tikzpicture}[baseline=(t1)]
\begin{feynman}[small]
    \vertex  (t1) {$t$}; 
    \vertex (4F) [square dot, scale=\sizesqdot, right = 25 pt of t1,color=red] {};
    \vertex (t2) [right= 25 pt of 4F] {$t$};
    \vertex  (inv) [scale = 0.01,above = 20 pt of 4F] {$t$};
    \diagram* {
    (t1)  -- [fermion] (4F) -- [fermion] (t2),
    (inv)  -- [fermion, half right] (4F) -- [fermion,half right] (inv),
}; 
\end{feynman}
\end{tikzpicture} \Biggl |_\text{FIN} &= 
\frac{1}{\Lambda^2}\left( \coeff{Qt}{(1)} + \frac{4}{3} \coeff{Qt}{(8)}\right) \\ 
& \times (B_{m_t}+K_{m_t}) \times 
\begin{tikzpicture}[baseline=(t1)]
\begin{feynman}[small]
    \vertex  (t1) [] {$t$}; 
    \vertex (4F) [dot, scale=0.01, right = 25 pt of t1,color=black] {};
    \vertex (t2) [right= 25 pt of 4F] {$t$};
    \node[shape=star,star points=4,star point ratio = 15,fill=black, draw,scale = 0.05, rotate=45] at (4F) {};
    \diagram* {
    (t1)  -- [fermion] (4F) -- [fermion] (t2),
     };
\end{feynman}
\end{tikzpicture} ,
\label{eq:diag_mt}
\end{aligned}
 \end{equation}

\begin{equation}
K_{m_t} = 
\begin{cases}
- \frac{m_t^2}{8\pi^2} & \text{(NDR)}\\
0 & \text{(BMHV)}.
\end{cases}
\label{eq:Kmt}
\end{equation}
Also in this case, $B_{m_t}$ denotes the scheme-independent contribution
\begin{equation} 
B_{m_t} = {m_t^2} \times \frac{ \log \left(\frac{\tilde{\mu}^2 }{m_t^2}\right)+1}{4 \pi ^2}.
\end{equation}
The results in Eqs.~(\ref{eq:diag_tg}, \ref{eq:diag_ghtt}, \ref{eq:diag_mt}) deserve some discussion. Equation~(\ref{eq:diag_tg}) shows that the chromomagnetic and four-top operators are closely linked, even though the latter operators come with an explicit loop diagram. A possible interpretation of this fact is that, under the assumption that the UV-complete theory is renormalisable and that the SM fields are weakly coupled to the unknown fields,\footnote{In presence of strongly coupled and/or non-renormalisable UV completions, operators as $\mathcal{O}_{\phi B} \equiv \left( \phi^\dagger \phi \right) B^{\mu \nu}B_{\mu \nu}$, which are expected to be generated at loop level in weakly coupled theories, can be generated at tree-level \cite{Jenkins:2013fya}.} there are operators which cannot be generated at tree-level. 
This means that their Wilson coefficients are expected to contain a loop suppression factor $1/(4\pi)^2$~\cite{Arzt:1994gp,Buchalla:2022vjp}. 
The power counting can be formalised conveniently via the chiral dimension $d_{\chi}$, supplementing the canonical dimension counting in $1/\Lambda$. 
We should mention that the chiral counting is highly non-trivial and cannot be made without some assumptions on the UV completion or kinematic regime, see Ref.~\cite{Brivio:2017vri} for an in-depth discussion. 
However, accepting the above-mentioned minimal assumptions, 
the tree-level diagram associated with the (loop-generated) operator $\mathcal{O}_{\phi G}$ enters the $gg\to h$ amplitude at the same power as the (tree-generated) operator $\mathcal{O}_{t\phi }$ inserted into a SM-like loop diagram, which is $1/(4\pi)^2\,1/\Lambda^2$. Similarly, $\mathcal{O}_{t G}$ inserted into a  one-loop diagram for $gg\to h$ (see Fig.~\ref{fig:cm_1loop}) and the two-loop diagram stemming from the insertion of the four-top operators into the $gg\to h$ matrix element (Fig.~\ref{fig:4t_2loop_gtt}) are of the same power, which is $1/(4\pi)^4\,1/\Lambda^2$. In the former case a loop-generated operator is inserted into a one-loop diagram, while in the latter case a tree-generated operator is contained in an explicit two-loop diagram. 
Therefore, in Eq.~\eqref{eq:Lag2ts}, $\coeff{tG}{}$ contains a loop suppression factor $1/(4\pi)^2$ relative to $\coeff{t\phi}{}$, the same holds for $\coeff{\phi G}{}$. Equation~\eqref{eq:diag_ghtt} shows that $g_{h\bar{t}t}$ ($\coeff{t \phi}{}$) and the four-top operators are also linked, however this relation comes with a relative suppression factor $1/\Lambda^2 \times 1/(4\pi)^2 $ ($1/(4\pi)^2 $). We remark that the connection between $g_{h \bar{t}t}$ and $\coeff{t \phi}{}$ is given by Eq.~\eqref{eq:ghttmtBF}.

\section{\label{sec:calculation} Calculation of the Higgs-gluon coupling}
In this section, we compute the four-top operator contribution at two-loop order to the Higgs-gluon coupling in the two different $\gamma_5$ schemes introduced in Sec.~\ref{sec:gamma5}. In the previous section we have shown that this contribution cannot be separated from that of the operators of $\mathcal{L}_{\text{2t}}$ in Eq.~\eqref{eq:Lag2ts} (cfr.~Eq.\eqref{eq:ghttmtBF}). In the case of $gg \to h$, we express the renormalised amplitude as follows
\begin{equation}
\mathcal{M}_{\text{TOT}} = \mathcal{M}_{\text{SM}}+\mathcal{M}_{\text{EFT}}, 
\end{equation}
\begin{equation}
\begin{aligned}
  \mathcal{M}_\text{EFT} = \frac{1}{\Lambda^2} \{ &
  \coeff{4t}{} \mathcal{M}_{4t} 
  + \coeff{tG}{} \mathcal{M}_{tG} +  \coeff{t \phi}{}\mathcal{M}_{t \phi } \\ 
  +& \coeff{\phi G}{}\mathcal{M}_{\phi G}
  + \mathcal{M}_{ \text{C.T.} }\}, 
\end{aligned}
\end{equation}
where $\mathcal{M}_{\mathrm{4t}}$ denotes the two-loop contribution of four-top operators and $\mathcal{M}_{tG}$ ($\mathcal{M}_{t \phi}$) the one-loop contribution of $\mathcal{O}_{tG}$ ($\mathcal{O}_{t\phi}$). 
The inclusion of $\mathcal{O}_{\phi G}$ is required in order to cancel the divergent part coming from $\mathcal{M}_{tG}$. $\mathcal{M}_{\phi G}$ represents its tree-level insertion, namely $-4 v \delta^{A_1 A_2} L^{\mu_1 \mu_2}$,
with
\begin{equation} \label{eq:Lorentz}
L^{\mu_1 \mu_2} = (m_h^2/2   ~g^{\mu_1 \mu_2}- p_1^{\mu_2}p_2^{\mu_1} ),
\end{equation}
and $A_1,A_2$ being the colour indices of the gluons.

The contribution from $\mathcal{O}_{t \phi}$ manifests itself as a modification of the SM parameters $g_{h \bar{t}t}^{\mathrm{SM}},\,m_t^{\mathrm{SM}}$ (see Eq.~\eqref{eq:ghttmtBF}). We can write, at $\order{1/\Lambda^2}$, 
\begin{equation}
\mathcal{M}_{\text{SM}}(g_{h \bar{t}t}^{\mathrm{SM}},m_t^{\mathrm{SM}}) + \frac{\coeff{t \phi}{}}{\Lambda^2} \mathcal{M}_{t \phi} \equiv \mathcal{M}_{\text{SM}}(g_{h \bar{t}t},m_t).
\end{equation}
In the following, we thus consider such SMEFT contribution to be included in the SM amplitude, provided that $g_{h \bar{t}t},m_t$ are given by Eq.~\eqref{eq:ghttmtBF}.

The four-top contribution to $\mathcal{M}_\text{EFT}$ can be split according to the different topologies of the associated Feynman diagrams. In Fig.~\ref{fig:4t_2loop} we show a sample of the 12 diagrams that need to be computed. The first topology is related to a correction to the Higgs-top-quark coupling (\ref{fig:4t_2loop_yt}), the second one to a correction to the top quark propagator (\ref{fig:4t_2loop_mt}) and the third one to a correction to the gluon-top vertex (\ref{fig:4t_2loop_gtt}). We group the first two classes of diagrams in $\mathcal{A}_{g_{h \bar{t}t}+m_t}$ (whose 
expression is discussed in Sec.~\ref{subsec:renormalisation})
and the third one in $\mathcal{A}_{g\bar{t}t}$.
We generated the diagrams with {\tt qgraf-3.6.5} \cite{Nogueira:1991ex} and performed the algebra with {\tt FeynCalc} \cite{Mertig:1990an, Shtabovenko:2016sxi,Shtabovenko:2020gxv}. Following the above classification, we express the four-top contribution as
\begin{equation}\label{eq:M4t}
\begin{aligned}
\coeff{4t}{}\mathcal{M}_{4t} &=  \mathcal{A}_{g_{h \bar{t}t}+m_t}  \left( \coeff{Qt}{(1)} + \frac{4}{3} \coeff{Qt}{(8)}\right) \frac{1}{\Lambda^2} \\
&+
\mathcal{A}_{g\bar{t}t} \left( \coeff{Qt}{(1)} - \frac{1}{6} \coeff{Qt}{(8)}\right) \frac{1}{\Lambda^2}.
\end{aligned}
\end{equation}
We note that the contribution from the operators $\mathcal{O}_{QQ}^{(1)},\,\mathcal{O}_{QQ}^{(3)},\,\mathcal{O}_{tt}$ vanishes in both schemes.
The two different combinations of the Wilson coefficients in Eq.~\eqref{eq:M4t} arise from the colour algebra.
We find that the result of $ \mathcal{A}_{g\bar{t}t}$ can be expressed in terms of the contribution to the amplitude due to an insertion of the chromomagnetic operator
\begin{equation} \label{eq:Agtt}
    \mathcal{A}_{g\bar{t}t} = 
    \left[ \frac{1}{2} K_{tG} \mathcal{M}_{tG}|_\text{DIV} + K_{tG}  \mathcal{M}_{tG}|_\text{FIN} \right],
\end{equation}
where $K_{tG}$ is the same as in Eq.~\eqref{eq:Ktg}. The divergent ($\mathcal{M}_{tG}|_\text{DIV}$) and finite ($\mathcal{M}_{tG}|_\text{FIN}$) parts of $\mathcal{M}_{tG}$ are given, respectively, by
\begin{equation}
   \mathcal{M}_{tG}|_\text{DIV} = - g_s m_t ~\frac{1}{{\epsilon}}  \frac{\sqrt{2}}{ 2 \pi^2} L^{\mu_1 \mu_2}  \epsilon_{\mu_1}(p_1) \epsilon_{\mu_2} (p_2) \delta^{A_1A_2},
\end{equation}
\begin{equation}
    \begin{aligned}
\mathcal{M}_{tG}|_\text{FIN} &= - \frac{g_s m_t \sqrt{2}}{4 \pi^2} L^{\mu_1 \mu_2}  \epsilon_{\mu_1}(p_1) \epsilon_{\mu_2} (p_2) \delta^{A_1A_2} \\
       & \times \Biggl(\frac{1}{4} \tau \log^2 \left(\frac{\beta-1}{\beta + 1} \right) + \beta \log\left(\frac{\beta-1}{\beta + 1} \right) \\
       &+ 2 \log\left(\frac{\tilde{\mu}^2}{m_t^2} \right) + 1 \Biggl).
    \end{aligned}
\end{equation}

We point out that the fact that $K_{tG}$ factorises in Eq.~\eqref{eq:Agtt} does not depend on the scheme.
The value of $K_{tG}$ depends on the scheme, and in particular $K_{tG}=0$ in BMHV. Remarkably, this implies that the structure of the divergences is different between the two schemes. This happens because of the combination of a scheme-independent pole of a loop integral with the scheme-dependent finite terms in Eq.~\eqref{eq:diag_tg}.
On the other hand, we find that the  divergent terms in $\mathcal{A}_{g_{h \bar{t}t}+m_t}$ are scheme-independent.

\begin{figure*}[!ht]
\centering
    \begin{subfigure}{0.3\textwidth}
        \centering
        \begin{tikzpicture} 
            \begin{feynman}[small]
                \vertex  (g1)  {$g$};
                \vertex  (gtt1) [dot,scale=\sizedot,right = of g1] {};
                \vertex (4F) [square dot, scale=\sizesqdot,below right= of gtt1,color=red] {};
                \vertex  (gtt2) [dot,scale=\sizedot,below left = of 4F] {};
                \vertex  (g2) [left=of gtt2]  {$g$};
                \vertex (htt) [dot,scale=\sizedot,right = 20 pt  of 4F] {};
                \vertex (h) [right = of htt] {$h$};

                \diagram* {
                    (g1)  -- [gluon] (gtt1),
                    (g2) -- [gluon] (gtt2),
                    (h)  -- [scalar] (htt),
                    (gtt1) -- [fermion] (4F) 
                    -- [fermion] (gtt2)
                     -- [fermion] (gtt1), 
                     (4F) -- [fermion, half left] (htt) -- [fermion, half left] (4F)
                };
            \end{feynman}
        \end{tikzpicture}
        \caption{Contribution to the Higgs-top quark coupling.}\label{fig:4t_2loop_yt}
    \end{subfigure}
\begin{subfigure}{0.3\textwidth}
\centering
\begin{tikzpicture} 
            \begin{feynman}[small]
                \vertex  (g1)  {$g$};
                \vertex  (gtt1) [dot,scale=\sizedot,right = of g1] {};
                \vertex (htt) [dot,scale=\sizedot,below right = of gtt1] {};
                \vertex (4F) [square dot, scale = \sizesqdot,color=red,below right= 14 pt of gtt1] {};
                \vertex (inv1) [scale=0.01,above right = 18 pt of 4F] {};
                \vertex  (gtt2) [dot,scale=\sizedot,below left = of htt] {};
                \vertex  (g2) [left=of gtt2]  {$g$};
                \vertex (h) [right = of htt] {$h$};
                \diagram* {
                    (g1)  -- [gluon] (gtt1),
                    (g2) -- [gluon] (gtt2),
                    (h)  -- [scalar] (htt),
                    (gtt1) -- [fermion] (4F) -- [fermion] (htt) 
                    -- [fermion] (gtt2)
                     -- [fermion] (gtt1), 
                     (4F) -- [fermion, half left] (inv1) -- [fermion, half left] (4F)
};
            \end{feynman}
        \end{tikzpicture}
        \caption{Contribution to the top quark propagator.}\label{fig:4t_2loop_mt}
    \end{subfigure}
    \begin{subfigure}{0.3\textwidth}
        \centering
        \begin{tikzpicture} 
            \begin{feynman}[small]
                \vertex  (g1)  {$g$};
                 \vertex (gtt1) [dot,scale=\sizedot,right= 28 pt of g1] {};
                \vertex  (4F) [square dot, scale=\sizesqdot, right = 20 pt of gtt1,color=red] {};
                \vertex (htt) [dot,scale=\sizedot,below right = of 4F] {};
                \vertex  (gtt2) [dot,scale=\sizedot,below left = of htt] {};
                \vertex  (g2) [left= 50 pt of gtt2]  {$g$};
                \vertex (h) [right = of htt] {$h$};

                \diagram* {
                    (g1)  -- [gluon] (gtt1),
                    (g2) -- [gluon] (gtt2),
                    (h)  -- [scalar] (htt),
                    (gtt1) -- [fermion, half right] (4F) 
                    -- [fermion,half right] (gtt1),
                     (4F) -- [fermion] (htt) -- [fermion] (gtt2) -- [fermion] (4F)
                };
            \end{feynman}
        \end{tikzpicture}
        \caption{Contribution to the gluon-top quark vertex.}\label{fig:4t_2loop_gtt}
    \end{subfigure}
\caption{Contributions from insertions of four-top quark operators (red square dot) to $gg \to h$ at two-loop level.} \label{fig:4t_2loop}
\end{figure*}
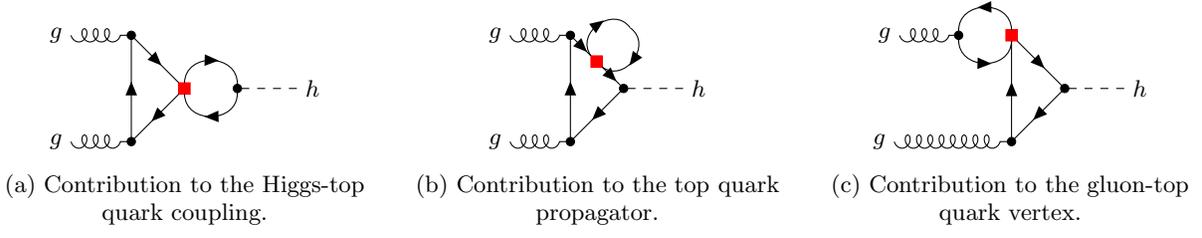

\subsection{Renormalisation} \label{subsec:renormalisation}
We use the minimal subtraction (${\text{MS}}$) renormalisation prescription for all the parameters in the theory. Schematically, the counterterms needed to renormalise the amplitude are given by
\begin{equation}
\label{eq:counterterm_diag}
\mathcal{M}_{\text{C.T.}}=
\begin{tikzpicture}[baseline=(htt)] 
            \begin{feynman}
                \vertex  (g1)  {$g$};
                 \vertex (gtt1) [dot,scale=\sizedot,right= 25 pt of g1] {};
                \vertex (htt) [dot,scale=\sizedot,below right = 20 pt of gtt1] {};
                \vertex  (gtt2) [dot,scale=\sizedot,below left = 20 pt of htt] {};
                \vertex  (g2) [left= 25 pt of gtt2]  {$g$};
                \vertex (h) [right = 20 pt of htt] {$h$};
               \node[shape=star,star points=5,star point ratio = 2,fill=black, draw,scale = 0.4] at (htt) {};
                \diagram* {
                    (g1)  -- [gluon] (gtt1),
                    (g2) -- [gluon] (gtt2),
                    (h)  -- [scalar] (htt),
                     (gtt1) -- [fermion] (htt) -- [fermion] (gtt2) -- [fermion] (gtt1)
                };
            \end{feynman}
        \end{tikzpicture}+
        \begin{tikzpicture}[baseline=(htt)] 
            \begin{feynman}
                \vertex  (g1)  {$g$};
                 \vertex (gtt1) [dot,scale=\sizedot,right= 25 pt of g1] {};
 \vertex (ct) [dot,scale=\sizedot,below right = 10. pt of gtt1] {};                \vertex (htt) [dot,scale=\sizedot,below right = 20 pt of gtt1] {};
                \vertex  (gtt2) [dot,scale=\sizedot,below left = 20 pt of htt] {};
                \vertex  (g2) [left= 25 pt of gtt2]  {$g$};
                \vertex (h) [right = 20 pt of htt] {$h$};
               \node[shape=star,star points=5,star point ratio = 2,fill=black, draw,scale = 0.4] at (ct) {};
                \diagram* {
                    (g1)  -- [gluon] (gtt1),
                    (g2) -- [gluon] (gtt2),
                    (h)  -- [scalar] (htt),
                     (gtt1) -- [fermion] (htt) -- [fermion] (gtt2) -- [fermion] (gtt1)
                };
            \end{feynman}
        \end{tikzpicture}+
          \begin{tikzpicture}[baseline=(hgg)] 
            \begin{feynman}
                \vertex  (g1)  {$g$};
                \vertex (hgg) [dot,scale=\sizedot,below right = 27 pt of g1] {};
                \vertex  (g2) [below left= 27 pt of hgg]  {$g$};
                \vertex (h) [right = 20 pt of hgg] {$h$};
               \node[shape=star,star points=5,star point ratio = 2,fill=black, draw,scale = 0.4] at (hgg) {};
                \diagram* {
                    (g1)  -- [gluon] (hgg),
                    (g2) -- [gluon] (hgg),
                    (h)  -- [scalar] (hgg),
                };
            \end{feynman}
        \end{tikzpicture}.
\end{equation}
For the top quark mass we have
\begin{equation}
m_t^{{\text{MS}}}=m_t^{(0)}+\delta m_t,
\end{equation}
with
\begin{equation}\label{eq:MSBarmtCT}
\delta m_t =\frac{m_t^3}{4 \pi^2 \Lambda^2 \, {\epsilon}} \left( \coeff{Qt}{(1)}+ \frac{4}{3} \coeff{Qt}{(8)} \right).
\end{equation} 
We note that typically in the computation of $gg\to h$ the top quark mass is renormalised in the on-shell scheme. In order to simplify our point (as we find the same ${\text{MS}}$ counterterm in NDR and BMHV) we restrict the discussion here to a pure ${\text{MS}}$ renormalisation.

In addition, the Wilson coefficient  $\coeff{t\phi}{}$, which mixes with the four-top operators via renormalisation group equation (RGE) running, needs to be renormalised. The coefficient of the operator is renormalised according to
\begin{equation} \label{eq:Ctphict}
\coeff{t\phi}{{\text{MS}}}=\coeff{t\phi}{(0)} +\delta \coeff{t \phi}{} \quad \text{   with   }\quad \delta \coeff{t \phi}{} = -\frac{1}{2 {\epsilon}}\frac{1}{16 \pi^2} \gamma_{t\phi,j} \coeff{j}{}\,,
\end{equation}
where $\gamma$ denotes the one-loop anomalous dimension of the SMEFT. The entries relevant for our discussion can be obtained from Refs.~\cite{rge1, rge2}. The equation correlating $\delta \coeff{t \phi}{}$ and the anomalous dimension matrix in Eq.~\eqref{eq:Ctphict} is discussed in detail in App.~\ref{app:RGE}.
The only four-top Wilson coefficients contributing to $\gamma_{t\phi,j} \coeff{j}{}$ are $\coeff{Qt}{(1,8)}$. The operator 
$\mathcal{O}_{t \phi}$ modifies the Higgs couplings to top quarks as discussed previously, see Eq.~\eqref{eq:ghttmtBF}. 

In analogy to $m_t$, we have:
\begin{equation}
g_{h \bar{t} t}^{{\text{MS}}}=g_{h \bar{t} t}^{(0)}+\delta g_{h \bar{t} t},
\end{equation}
with
\begin{equation}\label{eq:MSBarghttCT}
\delta g_{h \bar{t} t}=g_{h \bar{t}t} \frac{\left( 6 m_t ^2 - m_h^2 \right)}{8 \pi^2 \Lambda^2 \, {\epsilon}} \left( \coeff{Qt}{(1)}+\frac{4}{3} \coeff{Qt}{(8)} \right).
\end{equation}
From now on we will drop the superscript ${\text{MS}}$, leaving understood that all the parameters are renormalised in the ${\text{MS}}$ scheme. We recall that the divergent parts of the diagrams in Figs.~\ref{fig:4t_2loop_yt} and \ref{fig:4t_2loop_mt} are equal in the NDR and BMHV schemes, and they are fully removed by one-loop diagrams with an insertion of the one-loop counterterms in Eqs.~\eqref{eq:MSBarmtCT}, \eqref{eq:MSBarghttCT}.

The insertion of the chromomagnetic operator (see Fig.~\ref{fig:cm_1loop}) gives a divergent contribution to the Higgs-gluon coupling at one loop \cite{rge2, Grazzini:2018eyk,Deutschmann:2017qum}. We find this contribution to be scheme-independent. To remove all the divergences we need to choose (see Eq.~\eqref{eq:Agtt})
\begin{equation}\label{eq:ctCphiG}
\delta_{\phi G} = \frac{ g_{h\bar{t}t} g_s}{\Lambda^2 {\epsilon} \, 4 \sqrt{2} \pi^2} \left( \coeff{tG}{} + \frac{K_{tG}}{2} \left(\coeff{Qt}{(1)} - \frac{1}{6} \coeff{Qt}{(8)} \right) \right).
\end{equation}
This entails an important consequence: the anomalous dimension is scheme-dependent, as it contains the scheme-dependent $K_{tG}$. From $d \coeff{\phi G}{(0)}/ d \mu=0$, we obtain
\begin{equation}
16 \pi^2 \mu \frac{d \coeff{\phi G}{}}{d \mu} = -4 \sqrt{2} g_{h \bar{t}t} g_s \left( \coeff{tG}{} + K_{tG} \left(\coeff{Qt}{(1)} - \frac{1}{6} \coeff{Qt}{(8)} \right) \right).
\label{eq:RGECphiG}
\end{equation}
Notice that there is a relative factor of 2  between the contributions from $\coeff{Qt}{(1,8)}$ in Eq.~\eqref{eq:ctCphiG} and Eq.~\eqref{eq:RGECphiG}. This is a consequence of the contribution proportional to $
\coeff{tG}{}$ being $\order{g_{h \bar{t}t} g_s}$ and the contribution proportional to $\coeff{Qt}{(1,8)}$ being $\order{g_{h \bar{t}t}^2 g_s^2}$.\footnote{Using $g_{h \bar{t}t} = m_t/v + \order{1/\Lambda^2}$.} This (merely algebraic) fact will have important consequences, as we will show in the following. The details can be found in App.~\ref{app:RGE}.
We stress that the form of the RGE in Eq.~\eqref{eq:RGECphiG} shows that the contributions of $\coeff{tG}{}$, $\coeff{Qt}{(1,8)}$ enter at different loop orders (being $K_{tG} = \order{1/(4 \pi)^2}$). However, when the loop counting from Ref.~\cite{Buchalla:2022vjp} is considered, they enter at the same order, as explained in Sec.~\ref{sec:preliminary}. 

The differences in NDR and BMHV originating from the finite mixing of the four-fermion operators with chiral structure $(\bar{L}{L})(\bar{R}{R})$ into the chromomagnetic operator are well known, in particular in the context of flavour physics. This effect can induce a scheme-dependent anomalous dimension matrix at leading order \cite{Ciuchini:1993fk,Ciuchini:1993ks,Buras:1993xp, Herrlich:1994kh,Dugan:1990df}.
Using the strategy proposed in \cite{Ciuchini:1993fk,Ciuchini:1993vr, Buras:1993xp}, we can perform a finite renormalisation of the chromomagnetic operator and write 
\begin{equation}
\label{eq:finrenCM}
\coeff{tG}{} \to \coeff{tG}{} + K_{tG} \left(\coeff{Qt}{(1)} - \frac{1}{6} \coeff{Qt}{(8)} \right).
\end{equation}
This choice ensures a scheme-independent anomalous dimension matrix.

\subsection{Renormalised amplitude}
In the previous section we discussed how to obtain the same anomalous dimension matrix in both schemes. This is achieved via the inclusion of the effects of a scheme-dependent finite mixing in the Wilson coefficients. These effects are related to one-loop subdiagrams as in Eq.~\eqref{eq:diag_tg}.
One may wonder if redefinitions similar to Eq.~\eqref{eq:finrenCM} are enough to obtain the same result for the finite part of the amplitude in both schemes. In other words, we want to check if the scheme-dependence of the two-loop amplitude can be accounted for simply by computing one-loop subdiagrams.
The only scheme-dependent terms in the amplitudes are the ones stemming from a two-loop insertion of the four-top operators and they are parametrised by $K_{tG}$, $K_{g_{h\bar{t}t}}$ and $K_{m_t}$.  

We express the renormalised contribution from the diagrams in Figs.~\ref{fig:4t_2loop_yt},~\ref{fig:4t_2loop_mt} as \\ 
\begin{equation} \label{eq:Amtghtt}
\begin{aligned}
\mathcal{A}_{g_{h \bar{t}t} + m_t}^{\text{Ren}} &= \mathcal{M}_{g_{h \bar{t}t} + m_t}^{\text{S.I.}}\\
& + K_{g_{h \bar{t}t}} \mathcal{M}^{\text{SM}} + K_{m_t} \frac{\partial\mathcal{M}^{\text{SM}}}{\partial m_t } \times m_t,
\end{aligned}
\end{equation}
where $\mathcal{M}^{\text{SM}},\,\mathcal{M}_{g_{h \bar{t}t} + m_t}^{\text{S.I.}}$ are scheme-independent and they can be found in App.~\ref{app:results}. 
Putting together Eqs.~\eqref{eq:M4t},\eqref{eq:Agtt} and \eqref{eq:Amtghtt}
we have the following expression for the renormalised matrix element
\begin{equation}
\begin{aligned}
\mathcal{M}_\text{TOT}^{\text{Ren}} =& \left( \coeff{Qt}{(1)} + \frac{4}{3} \coeff{Qt}{(8)}\right) \frac{1}{\Lambda^2} \mathcal{M}_{g_{h \bar{t}t} + m_t}^{\text{S.I.}}  \\
+&\left[  \coeff{tG}{} + 
\left( \coeff{Qt}{(1)} - \frac{1}{6} \coeff{Qt}{(8)}\right)
 K_{tG}   \right]  \frac{1}{\Lambda^2} \mathcal{M}_{tG}|_\text{FIN} \\ 
+& \left[ 1+ \left( \coeff{Qt}{(1)} + \frac{4}{3} \coeff{Qt}{(8)}\right) \frac{1}{\Lambda^2} K_{h \bar{t}t} 
\right]\mathcal{M}_{\text{SM}}
\\ 
+&\left( \coeff{Qt}{(1)} + \frac{4}{3} \coeff{Qt}{(8)}\right) \frac{1}{\Lambda^2}   K_{m_t} \frac{\partial \mathcal{M}_{\text{SM}}}{\partial m_t}  \times m_t \\
+& \; \coeff{\phi G}{} \mathcal{M}_{\phi G} \frac{1}{\Lambda^2} .
\end{aligned}
\end{equation}
We note that $\mathcal{M}_\text{TOT}^{\text{Ren}}$ represents a physical on-shell scattering amplitude, which must be scheme-independent.\footnote{This can be best understood from a top-down perspective.} Therefore, the scheme-dependence of the $K$-terms has to be compensated by a scheme-dependence of the parameters. To make this more evident, we define the following set of parameters identified by a tilde
\begin{align}
\tilde{\mathcal{C}}_{tG}  &=\coeff{tG}{} +  \left( \coeff{Qt}{(1)} - \frac{1}{6} \coeff{Qt}{(8)}\right)
 K_{tG}, \label{eq:CtGTilde}\\
 \tilde{g}_{h \bar{t}t}  &={g}_{h \bar{t}t}\left[1+
 \left( \coeff{Qt}{(1)} + \frac{4}{3} \coeff{Qt}{(8)}\right) \frac{1}{\Lambda^2} K_{h \bar{t}t} 
 \right],  \label{eq:ghttTilde} \\
 \tilde{m}_{t}  &={m}_{t}\left[1+
 \left( \coeff{Qt}{(1)} + \frac{4}{3} \coeff{Qt}{(8)}\right) \frac{1}{\Lambda^2} K_{m} 
 \right]. \label{eq:mtTilde}
\end{align}
Noting that, under a redefinition of the top mass $m_t \to m_t + \Delta m_t$, one has $\mathcal{M}_{\text{SM}} \to \mathcal{M}_{\text{SM}} +\Delta m_t \partial \mathcal{M}_{\text{SM}}/ \partial m_t $,
we can write the total matrix element in a more compact form (at $\order{1/\Lambda^2}$):
\begin{equation}
\begin{aligned}
\mathcal{M}_\text{TOT}^{\text{Ren}} =& \left( \coeff{Qt}{(1)} + \frac{4}{3} \coeff{Qt}{(8)}\right) \frac{1}{\Lambda^2} \mathcal{M}_{g_{h \bar{t}t} + m_t}^{\text{S.I.}} \\
+&  \frac{\tilde{\mathcal{C}}_{tG}}{\Lambda^2} \mathcal{M}_{tG}|_\text{FIN}+ \mathcal{M}_{\text{SM}}( \tilde{g}_{h \bar{t}t}, \tilde{m}_{t})
+ \; \frac{\coeff{\phi G}{}}{\Lambda^2}  \mathcal{M}_{\phi G} .
\end{aligned}
\end{equation}
In the previous expression, $\mathcal{M}_{\text{SM}}( \tilde{g}_{h \bar{t}t}, \,\tilde{m}_{t})$ is given by Eq.~\eqref{eq:MSM} where $g_{h \bar{t}t},\, m_t$ are replaced by $\tilde{g}_{h \bar{t}t}, \tilde{m}_t$.
From the amplitudes $\mathcal{M}_{g_{h \bar{t}t} + m_t}^{\text{S.I.}}, \, \mathcal{M}_{tG}, \, \mathcal{M}_{\text{SM}}, \, \mathcal{M}_{\phi G}$ being scheme-independent, it follows that the  combinations in Eqs.~(\ref{eq:CtGTilde}-\ref{eq:mtTilde}) must be scheme-independent. 

It should be stressed that Eq.~\eqref{eq:CtGTilde} is the same relation we obtained in the previous section, namely Eq.~\eqref{eq:finrenCM}: the same finite shift makes both the anomalous dimension matrix and the renormalised amplitude scheme-independent. 
We also remark that, at the order we are working, $g_{h \bar{t}t}$ and $m_t$ can be used interchangeably with $\tilde{g}_{h \bar{t}t}$ and $\tilde{m}_{t}$ in $\mathcal{M}_{tG,\phi G},\mathcal{M}_{g_{h \bar{t}t} + m_t}^{\text{S.I.}}$ because their contribution to $\mathcal{M}_{\text{TOT}}$ is already suppressed by $\order{1/\Lambda^2}$.

\subsection{Summary of the computation}
We can now summarise the differences between the two schemes. From Eqs.~(\ref{eq:CtGTilde}-\ref{eq:mtTilde}) it is evident that there exists a difference between the parameters in the two schemes which is proportional to $K_X^{\text{NDR}}-K_X^{\text{BMHV}}$. This quantity does not depend on the prescription used to identify the $K$-terms. 

In BMHV all the $K$-terms are vanishing, so the previous redefinitions are trivial. The scheme-independence condition $\tilde{X}_i^{\text{NDR}}=\tilde{X}_i^{\text{BMHV}}$ (being $X=\coeff{tG}{},m_t, {g}_{h\bar{t}t}$) allows us to write at $\order{1/\Lambda^2}$\footnote{
If we had included the loop factor $1/(4 \pi)^2$ explicitly in the ${\cal C}_{tG}$-term in the Lagrangian Eq.~\eqref{eq:Lag2ts}, it would be manifest that the chromomagnetic and the four-top operators contribute at the same order in the chiral counting, because in this case the factor $1/(4 \pi)^2$ in Eq.~\eqref{eq:CtGMap} would be absent. } 

\begin{align}
\coeff{tG}{\text{NDR}}  &=\coeff{tG}{\text{BMHV}} - \left( \coeff{Qt}{(1)} - \frac{1}{6} \coeff{Qt}{(8)}\right)
\frac{\sqrt{2} {g}_{h\bar{t}t} g_s}{16 \pi^2} , \label{eq:CtGMap}\\
 {g}_{h \bar{t}t}^{\text{NDR}}  &={g}_{h\bar{t}t}^{\text{BMHV}}
 - {g}_{h \bar{t}t} \left( \coeff{Qt}{(1)} + \frac{4}{3} \coeff{Qt}{(8)}\right)  \frac{(m_h^2-6m_t^2)}{16 \pi^2 \Lambda^2} 
,  \label{eq:ghttMap} \\
 {m}_{t}^{\text{NDR}}  &={m}_{t}^{\text{BMHV}} +
  \left( \coeff{Qt}{(1)} + \frac{4}{3} \coeff{Qt}{(8)}\right)  \frac{{m}_{t}^3}{8\pi^2 \Lambda^2}.  \label{eq:mtMap} 
\end{align}
The map described by Eqs.~(\ref{eq:CtGMap}-\ref{eq:mtMap}), establishes a connection between the two schemes. When such relations are considered, the two schemes give the same anomalous dimension matrix and the same renormalised amplitude. 

The last two equations can be recasted in terms of $\yuk{t},\coeff{t \phi}{}$ by means of Eq.~\eqref{eq:ghttmtBF}: 
\begin{align}
 \yuk{t}^{\text{NDR}}  &=\yuk{t}^{\text{BMHV}}
 +  \left( \coeff{Qt}{(1)} + \frac{4}{3} \coeff{Qt}{(8)}\right)  \frac{\lambda v^2 \yuk{t}}{16 \pi^2 \Lambda^2} 
,  \label{eq:YtMap} \\
 \coeff{t \phi }{\text{NDR}}  &= \coeff{t \phi }{\text{BMHV}} +
 \sout{\yuk{t}}  \left( \coeff{Qt}{(1)} + \frac{4}{3} \coeff{Qt}{(8)}\right)  \frac{\yuk{t}(\lambda -\yuk{t}^2)}{8\pi^2 },  \label{eq:CtPhiMap} 
\end{align}
where $\lambda = m_h^2/(2 v^2)  + \order{1/\Lambda^2}$. 

\section{\label{sec:matching} Matching with UV-models}
As discussed in the previous section, the differences in the finite terms of the amplitude when using the NDR and the BMHV scheme can be absorbed by different definitions of the parameters $\coeff{tG}{}, g_{h \bar{t}t}$, and $ m_t$. 
In this section we perform the matching with concrete UV completions of the SM, in order to validate our EFT approach from a top-down point of view. 
The matching is performed in the unbroken phase (following the notation used in Ref.~\cite{deBlas:2017xtg}), in which $g_{h \bar{t}t}$ and $m_t$ can be traded more conveniently in favour of $\coeff{t \phi}{}$ and $\yuk{t}$. In the remainder of the section we will use a thicker fermion line to denote the iso-doublet $Q_L$ and a thinner fermion line to denote the iso-singlet $t_R$ in the Feynman diagrams. We write the ${SU(3)}_{C} \otimes {SU(2)}_{L}\otimes{U(1)}_{{\mathsf{y}}}$ quantum numbers as $({\mathrm{R}}_{C},{\mathrm{R}}_{L})_{\mathsf{y}}$, $\mathrm{R}$ being the representation in which the particle transforms and $\mathsf{y}$ its hypercharge.

\subsection{New scalar: $\Phi \sim (8,2)_{\frac{1}{2}}$ } 
\begin{figure}[!t]
    \centering
    \begin{subfigure}[t]{0.45\linewidth}
        \centering
        \begin{tikzpicture} 
            \begin{feynman}[small]
                \vertex  (gtt) [dot, scale = \sizedot] {};
                \vertex (gi) [left= of gtt] {\(  g \)};
                \vertex (s1) [dot, scale = \sizedot,below right= of gtt] {};
                \vertex  (htt) [dot,scale = \sizedot,below left= of s1] {};
                \vertex (hi) [left = of htt] { $\phi^\dagger $};
                 \vertex (s2) [dot, scale = \sizedot,right = of s1] {};
                \vertex (tR) [above right =of s2] {$t_R$};
                \vertex (QL) [below right =of s2] {$Q_L$};
                \diagram* {
                    (gi)  -- [gluon] (gtt),
                    (hi)  -- [scalar] (htt),
                    (gtt) -- [anti fermion, line width = 1.5 pt] (s1) 
                    -- [anti fermion, line width = 0.3 pt] (htt)
                     -- [anti fermion, line width = 1.5 pt] (gtt), 
                     (s1) -- [scalar, line width = 2.0 pt, edge label = $\Phi$] (s2),
                    (QL) -- [fermion, line width = 1.5 pt] (s2) -- 
                    [fermion, line width = 0.3 pt] (tR)
                };
            \end{feynman}
        \end{tikzpicture}
        \caption{}\label{}
    \end{subfigure}
      \begin{subfigure}[t]{0.45\linewidth}
        \centering
        \begin{tikzpicture} 
            \begin{feynman}[small]
                \vertex  (gtt) [dot, scale = \sizedot] {};
                \vertex (gi) [left= of gtt] {\(  g \)};
                \vertex (s1) [dot, scale = \sizedot,below right= of gtt] {};
                \vertex  (htt) [dot,scale = \sizedot,below left= of s1] {};
                \vertex (hi) [left = of htt] { $\phi^\dagger $};
                 \vertex (s2) [dot, scale = \sizedot,right = of s1] {};
                \vertex (tR) [above right =of s2] {$t_R$};
                \vertex (QL) [below right =of s2] {$Q_L$};
                \diagram* {
                    (gi)  -- [gluon] (htt),
                    (hi)  -- [scalar] (gtt),
                    (gtt) -- [anti fermion, line width = 1.5 pt] (s1) 
                    -- [anti fermion, line width = 0.3 pt] (htt)
                     -- [anti fermion, line width = 0.3 pt] (gtt), 
                     (s1) -- [scalar, line width = 2.0 pt, edge label = $\Phi$] (s2),
                    (QL) --[fermion, line width = 1.5 pt] (s2) -- 
                    [fermion, line width = .3 pt] (tR)
                };
            \end{feynman}
        \end{tikzpicture}
        \caption{}\label{}
    \end{subfigure}
\caption{One-loop diagrams contributing to the matching with the chromomagnetic operator.} \label{fig:scalar1LmatchingOctet}
    \end{figure}
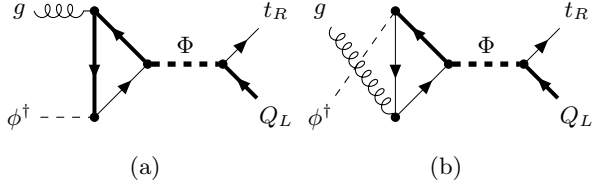

We consider, in addition to the SM, a new heavy scalar with a mass $M_{\Phi} \gg v$ and quantum numbers $\Phi \sim (8,2)_{\frac{1}{2}}$. The Lagrangian in this case can be written as
\begin{equation}
\begin{split}
\mathcal{L}_\Phi &= (D_\mu \Phi)^{\dagger}  D^\mu \Phi - M_\Phi^2 \Phi^\dagger \Phi \\ &- Y_{\Phi}  \left( \Phi^{A,\dagger} \varepsilon \bar{Q}_L ^T  T^A t_R + \hc \right),
\end{split}
\end{equation}
where $\varepsilon$ is the Levi-Civita pseudotensor in the isospin space and $T$ refers to the transposition in isospin space only. 
The tree-level matching yields 
\begin{equation}
\mathcal{L}=\frac{Y_{\Phi}^2}{ M_\Phi^2} (\bar{Q}_L T^A t_R) (\bar{t}_R T^A Q_L).
\end{equation}
This operator does not appear in the Warsaw basis since it is considered redundant in $D=4$ dimensions. In the following it will be referred to as $\mathcal{R}_{Qt}^{(8)}$. Using the Fierz identities, one can recast this result in terms of operators in the Warsaw basis \cite{deBlas:2017xtg}:
\begin{equation}\label{eq:4tmatchingOctet}
\frac{\coeff{Qt}{(1)}}{\Lambda^2} = -\frac{2}{9} \frac{Y_{\Phi}^2}{ M_\Phi^2}, \quad \frac{\coeff{Qt}{(8)}}{\Lambda^2} = \frac{1}{6} \frac{Y_{\Phi}^2}{ M_\Phi^2}.
\end{equation}
Now we compute the matching at one-loop level to the chromomagnetic operator. 
 The relevant diagrams are given in Fig.~\ref{fig:scalar1LmatchingOctet}, while diagrams with $t$-channel exchange within the loop are forbidden due to the conservation of hypercharge.

Evaluating the diagrams in Fig.~\ref{fig:scalar1LmatchingOctet} gives zero in both NDR and BMHV, in contrast with our previous observations. 
However, the Fierz identity we used for the matching of the four-fermion operators is broken by $\order{\epsilon}$ terms when dimensional regularisation is used ($D=4 -2 \epsilon$), as noted in Ref.~\cite{Fuentes-Martin:2022vvu}.  
Following this reference, we define the evanescent operator as 
\begin{equation}
\mathcal{E}=\mathcal{R}_{Qt}^{(8)} - \left(-\frac{2}{9}\mathcal{O}_{Qt}^{(1)}+\frac{1}{6}\mathcal{O}_{Qt}^{(8)} \right)
\end{equation}
and we compute its insertion (in both schemes). We find that in NDR the evanescent operator contributes to the matching to the chromomagnetic operator 
\begin{equation}
\begin{aligned}
 \mathrm{NDR} :\;  \frac{Y_{\Phi}^2}{ M_\Phi^2} \mathcal{R}_{Qt}^{(8)} &=  \overbrace{-\frac{2}{9}\frac{Y_{\Phi}^2}{M_\Phi^2}}^{\mathcal{C}_{Qt}^{(1)}/\Lambda^2} \mathcal{O}_{Qt}^{(1)}+\overbrace{\frac{1}{6}\frac{Y_{\Phi}^2}{M_\Phi^2}}^{\mathcal{C}_{Qt}^{(8)}/\Lambda^2}\mathcal{O}_{Qt}^{(8)} \\
 & \underbrace{
 +\frac{1}{16 \pi^2} \frac{\yuk{\Phi}^2}{ M_\Phi^2}\frac{g_s \yuk{t}}{4}}_{\mathcal{C}_{tG}/\Lambda^2} \mathcal{O}_{tG} + \hc\,.
\end{aligned}
\end{equation}
This result reproduces the term proportional to the chromomagnetic operator presented in Ref.~\cite{Fuentes-Martin:2022vvu}.\footnote{This reference uses a different convention for the covariant derivative with respect to the one used in Ref.~\cite{Dedes:2017zog}, which we follow in the Feynman rules. This leads to a relative minus sign in terms with an odd power of $g_s$. In addition, the different normalisation of the quartic Higgs self-coupling in Ref.~\cite{Fuentes-Martin:2022vvu} requires the replacements $\lambda/2 \to \lambda$, $\mu^2 \to \lambda v^2$  to convert their result into our conventions.}
In BMHV we obtain
\begin{equation}
\begin{split}
\mathrm{BMHV} :\; \frac{Y_{\Phi}^2}{ M_\Phi^2} \mathcal{R}_{Qt}^{(8)} &=  \overbrace{-\frac{2}{9}\frac{Y_{\Phi}^2}{M_\Phi^2}}^{\mathcal{C}_{Qt}^{(1)}/\Lambda^2} \mathcal{O}_{Qt}^{(1)}+\overbrace{\frac{1}{6}\frac{Y_{\Phi}^2}{M_\Phi^2}}^{\mathcal{C}_{Qt}^{(8)}/\Lambda^2}\mathcal{O}_{Qt}^{(8)}\,.
\end{split}
\end{equation}
We conclude that the difference between the NDR scheme and BMHV scheme (using Eq.~\eqref{eq:4tmatchingOctet} and $ \sqrt{2} \, m_t = \yuk{t}v +\order{1/\Lambda^2}$) is exactly the one described by Eq.~\eqref{eq:CtGMap}. 

Furthermore, we need to compute the matching to the top Yukawa coupling as well as to $ \coeff{t \phi}{}$. Doing so we find in both schemes zero, by colour. This is in trivial agreement with Eqs.~\eqref{eq:ghttMap},~\eqref{eq:mtMap} since, within this model, $\coeff{Qt}{(1)}+\frac{4}{3} \coeff{Qt}{(8)}=0$. In order to test Eqs.~\eqref{eq:ghttMap},~\eqref{eq:mtMap} we hence need to consider a different model, namely replacing the colour octet $\Phi$ with a colour singlet $\varphi$. 
\subsection{New scalar: $\varphi \sim (1,2)_{\frac{1}{2}}$ } 
We consider, in addition to the SM, a new heavy scalar with a mass $M_{\varphi} \gg v$ and quantum numbers $\varphi \sim (1,2)_{\frac{1}{2}}$. The Lagrangian in this case can be written as
\begin{equation}
\begin{split}
\mathcal{L}_\varphi &= (D_\mu \varphi)^{\dagger}  D^\mu \varphi - M_\varphi^2 \varphi^\dagger \varphi \\ &- Y_{\varphi}  \left( \varphi^{\dagger} \varepsilon \bar{Q}_L ^T   t_R + \hc \right).
\end{split}
\end{equation}
The tree-level matching yields 
\begin{equation}
\mathcal{L}=\frac{Y_{\varphi}^2}{ M_\varphi^2} (\bar{Q}_L t_R) (\bar{t}_R  Q_L).
\end{equation}
As in the previous case, this operator does not appear in the Warsaw basis being redundant in $D=4$ dimensions. In the following it will be referred to as $\mathcal{R}_{Qt}^{(1)}$. We find

\begin{equation}\label{eq:4tmatchingSinglet}
\frac{\coeff{Qt}{(1)}}{\Lambda^2} = -\frac{1}{6} \frac{Y_{\varphi}^2}{ M_\varphi^2}, \quad \frac{\coeff{Qt}{(8)}}{\Lambda^2} = - \frac{Y_{\varphi}^2}{ M_\varphi^2}.
\end{equation}
Due to colour structure, there are no contributions to the chromomagnetic operator. The tree-level matching implies $ \coeff{Qt}{(1)}-\frac{1}{6}\coeff{Qt}{(8)}=0$, in agreement with Eq.~\eqref{eq:CtGMap} since $\coeff{tG}{\text{NDR}}=\coeff{tG}{\text{BMHV}}=0$ within this model. \newline 
\begin{figure}[t]
    \centering
\begin{subfigure}[t]{0.45\linewidth}
    \begin{tikzpicture}
\begin{feynman}[small]
\vertex (htt) [dot, scale=\sizedot]  {};
\vertex (s1) [dot, scale=\sizedot, left=20 pt of htt]  {};
\vertex (s2) [dot, scale=\sizedot, left= 15 pt of s1]  {};
\vertex (hi) [right= 20 pt of htt] {$\phi^{\dagger}$};
\vertex (tR) [above left=of s2] {$t_R$};
\vertex (QL) [below left=of s2] {$Q_L$};
\diagram* {
(htt) -- [scalar] (hi),
(s1) -- [ fermion, line width=1.5pt, half left] (htt) -- [fermion, line width=0.3pt, half left] (s1),
(s2) -- [scalar, line width=2.0pt, edge label=$\varphi$] (s1),
(QL) -- [fermion, line width=1.5pt] (s2),
(s2) -- [fermion, line width=0.3pt] (tR),
};
\end{feynman}
\end{tikzpicture}
\caption{}\label{fig:scalar1LmatchingSinglet_yt}
\end{subfigure}
\begin{subfigure}[t]{0.45\linewidth}
    \centering
    \begin{tikzpicture}
\begin{feynman}[small]
\vertex (htt) [dot, scale=\sizedot]  {};
\vertex (htta) [dot, scale=\sizedot, above= 20 pt of htt] {};
\vertex (httb) [dot, scale=\sizedot, below= 20 pt of htt] {};
\vertex (hia) [right =20 pt of htta] {$\phi^{\dagger}$};
\vertex (hib) [right =20 pt of httb] {$\phi^{\dagger}$};

\vertex (s1) [dot, scale=\sizedot, left=20 pt of htt]  {};
\vertex (s2) [dot, scale=\sizedot, left= 15 pt of s1]  {};
\vertex (hi) [right=20 pt of htt] {$\phi$};
\vertex (tR) [above left=of s2] {$t_R$};
\vertex (QL) [below left=of s2] {$Q_L$};
\diagram* {
(htt) -- [scalar] (hi),
(htta) -- [scalar] (hia),(httb) -- [scalar] (hib),

(s1) -- [ fermion, line width=1.5pt] (htta) -- [fermion, line width=0.3pt] (htt) -- [ fermion, line width=1.5 pt] (httb) -- [ fermion, line width=0.3pt] (s1),
(s2) -- [scalar, line width=2.0pt, edge label=$\varphi$] (s1),
(QL) -- [fermion, line width=1.5pt] (s2),
(s2) -- [fermion, line width=0.3pt] (tR),
};
\end{feynman}
\end{tikzpicture}
\caption{}\label{fig:scalar1LmatchingSinglet_tH}

\end{subfigure}
\caption{One-loop diagrams contributing to the matching to the Yukawa coupling (left) and to $\coeff{t \phi} {}$ (right).} \label{fig:scalar1LmatchingSinglet}
\end{figure}
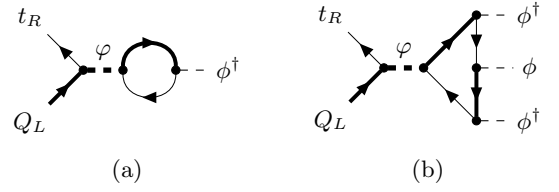
Following the procedure outlined in the previous section, we compute the diagrams in Fig.~\ref{fig:scalar1LmatchingSinglet} to obtain the contributions to $\yuk{t}$ and $\coeff{t\phi}{}$ in both schemes. The matching condition for $\yuk{t}$ ($\coeff{t\phi}{}$) is obtained subtracting from the diagram in Fig.~\ref{fig:scalar1LmatchingSinglet_yt} (\ref{fig:scalar1LmatchingSinglet_tH}) the one-loop amplitude for $\bar{Q}_L t_R \to \phi^{\dagger}$ ($\bar{Q}_L t_R \to \phi^{\dagger}\phi \phi^{\dagger}$) with an insertion of four-top operators. In other words, we are interested in computing the insertion of the evanescent operator:
\begin{equation}
\mathcal{E}=\mathcal{R}_{Qt}^{(1)} - \left( -\frac{1}{6}\mathcal{O}_{Qt}^{(1)}-\mathcal{O}_{Qt}^{(8)}  \right).
\end{equation}
In NDR we find:
\begin{equation}
\begin{aligned}
\mathrm{NDR} :\; \frac{Y_{\varphi}^2}{M_\varphi^2}  \mathcal{R}_{Qt}^{(1)} &=  \overbrace{-\frac{1}{6}\frac{Y_{\varphi}^2}{M_\varphi^2}}^{\mathcal{C}_{Qt}^{(1)}/\Lambda^2} \mathcal{O}_{Qt}^{(1)}\overbrace{-\frac{Y_{\varphi}^2}{M_\varphi^2}}^{\mathcal{C}_{Qt}^{(8)}/\Lambda^2}\mathcal{O}_{Qt}^{(8)} \\
 & \underbrace{+\frac{1}{16 \pi^2} \frac{\yuk{\varphi}^2}{ M_\varphi^2}\left(3 \yuk{t}^3 - 3 \lambda {\yuk{t}} \right)}_{\mathcal{C}_{t\phi}/\Lambda^2} \mathcal{O}_{t\phi} + \hc \\
 & \underbrace{-\frac{1}{16 \pi^2} \frac{\yuk{\varphi}^2}{ M_\varphi^2}\, \frac{3}{2} \lambda v^2  \yuk{t}}_{ \Delta \yuk{t}} \, \left(\bar{Q}_L \tilde{\phi} t_R \right)+ \hc 
 ,
\end{aligned}
\end{equation}
confirming once again the results obtained in Ref.~\cite{Fuentes-Martin:2022vvu}.  In this notation, $\Delta \yuk{t}$ represents the contribution to the top Yukawa coupling from the matching, while $\yuk{t}$ represents the coefficient of the four-dimensional Yukawa operator $\left(\bar{Q}_L \tilde{\phi} t_R \right)$.

In BMHV we find:
\begin{equation}
\mathrm{BMHV} :\; \frac{Y_{\varphi}^2}{M_\varphi^2}  \mathcal{R}_{Qt}^{(1)} =  \overbrace{-\frac{1}{6}\frac{Y_{\varphi}^2}{M_\varphi^2}}^{\mathcal{C}_{Qt}^{(1)}/\Lambda^2} \mathcal{O}_{Qt}^{(1)}\overbrace{-\frac{Y_{\varphi}^2}{M_\varphi^2}}^{\mathcal{C}_{Qt}^{(8)}/\Lambda^2}\mathcal{O}_{Qt}^{(8)} .
\end{equation}
Using Eq.~\eqref{eq:ghttmtBF} 
we can compute $m_t,\, g_{h t\bar{t}}$ and confirm Eqs.~\eqref{eq:ghttMap}, \eqref{eq:mtMap}. 

\section{\label{sec:moreinterplay}Interplay between more operators in the SMEFT}

The primary focus of this paper is the demonstration of $\gamma_5$ scheme differences in the treatment of four-top operators, since they provide a convenient playground for investigation due to the factorization of loop integrals.
However, considering a complete operator basis in SMEFT, there are other classes of operators that share similar features regarding the treatment of $\gamma_5$. 
Analogous to Sec.~\ref{sec:preliminary} (but more schematically) we demonstrate in the following that there is also a scheme-dependent finite mixing at one-loop order for operators in the class of $\psi^2\phi^2D$ of Ref.~\cite{dim6smeft}.

For the purpose of this discussion, we consider the two operators
\begin{equation}\label{eq:Lag2t2phi}
\begin{split}
{\cal L}_{2t2\phi}=&\frac{\coeff{\phi Q}{(1)}}{\Lambda^2}\bar{Q}_L\gamma_\mu Q_L\left(\phi^\dagger i\overleftrightarrow{D}^\mu\phi\right)
 \\ 
 +&\frac{\coeff{\phi t}{}}{\Lambda^2}\bar{t}_R\gamma_\mu t_R\left(\phi^\dagger i\overleftrightarrow{D}^\mu\phi\right)\;,
\end{split}
\end{equation}
where we introduced the short-hand notation
\begin{equation}
i\overleftrightarrow{D}^\mu=iD^\mu-i\overleftarrow{D}^\mu\;.
\end{equation}
Similar to the four-top operators in Eq.~\eqref{eq:Lag4t}, the operators in Eq.~\eqref{eq:Lag2t2phi} are composed of current-current interactions including chiral vector currents. 
These current-current operators can be generated by integrating out a new heavy vector particle at tree-level that couples to the SM currents. A concrete and comparably easy realization is given e.g.~by the Third Family Hypercharge Model \cite{Allanach:2018lvl,Allanach:2021bbd}.
We restrict the direct evaluation of one-loop contributions of the operators in Eq.~\eqref{eq:Lag2t2phi} to the gaugeless limit of the SM\footnote{
In the gaugeless limit, the SM gauge bosons are completely decoupled from the rest of the theory, taking the limit $g_1\to 0$ and $g_2\to 0$. 
The Goldstone fields of the SM Higgs doublet are therefore massless physical degrees of freedom.
The explicit analytic results in this section are equivalent to the pure Goldstone contribution in Landau gauge.
} and only investigate the contribution to the chromomagnetic form factor, since this is sufficient to point out the necessity of a more exhaustive study in future work.

An explicit evaluation of the one-loop correction to $g\to\bar{t}t$ in the broken phase leads to
\begin{equation}
\begin{aligned}
&\begin{tikzpicture}[baseline=(4F)]
            \begin{feynman}[small]
                \vertex  (g1)  {$g$};
                 \vertex (gtt1) [dot, scale=\sizedot, right= 20 pt of g1] {};
                \vertex  (gtG1) [square dot,scale=\sizesqdot,above right = 15 pt of gtt1, color = yellow] {};
                \vertex  (gtG2) [dot, scale=\sizedot,below right = 15 pt of gtt1] {};
                \vertex  (t1) [above right= 15 pt of gtG1] {$t$};
                \vertex (t2) [below right= 15 pt of gtG2] {$t$};
                \vertex  (l1) [right= 20 pt of gtt1] {$G^0$};

                \diagram* {
                    (g1)  -- [gluon] (gtt1),  
                    (t1) -- [anti fermion] (gtG1) -- [anti fermion] (gtt1) 
                    -- [anti fermion] (gtG2)-- [anti fermion] (t2),
                    (gtG1) -- [scalar] (gtG2)
                };
            \end{feynman}
        \end{tikzpicture}
        +\begin{tikzpicture}[baseline=(4F)]
            \begin{feynman}[small]
                \vertex  (g1)  {$g$};
                 \vertex (gtt1) [dot, scale=\sizedot, right= 20 pt of g1] {};
                \vertex  (gtG1) [dot, scale=\sizedot,above right = 15 pt of gtt1] {};
                \vertex  (gtG2) [square dot,scale=\sizesqdot,below right = 15 pt of gtt1, color = yellow] {};
                \vertex  (t1) [above right= 15 pt of gtG1] {$t$};
                \vertex (t2) [below right= 15 pt of gtG2] {$t$};
                \vertex  (l1) [right= 20 pt of gtt1] {$G^0$};

                \diagram* {
                    (g1)  -- [gluon] (gtt1),  
                    (t1) -- [anti fermion] (gtG1) -- [anti fermion] (gtt1) 
                    -- [anti fermion] (gtG2)-- [anti fermion] (t2),
                    (gtG1) -- [scalar] (gtG2)
                };
            \end{feynman}
        \end{tikzpicture}\Biggl |_\text{FIN}
        \\&\qquad= \frac{\coeff{\phi Q}{(1)}-\coeff{\phi t}{}}{\coeff{tG}{}} K_{tG}^{2t2\phi}  \times 
        \begin{tikzpicture}[baseline=(4F)]
            \begin{feynman}[small]
                \vertex  (g1)  {$g$};
                 \vertex (gtt1) [square dot, scale=\sizesqdot, right= 20 pt of g1, color=blue] {};
                \vertex  (t1) [above right= 25 pt of gtt1] {$t$};
                \vertex (t2) [below right= 25 pt of gtt1] {$t$};

                \diagram* {
                    (g1)  -- [gluon] (gtt1),  
                    (t2) -- [fermion] (gtt1) -- [fermion] (t1),};
            \end{feynman}
        \end{tikzpicture}
+\dots
\end{aligned}
        \label{eq:diag_tg_2t2phi}
\end{equation}
where the gluon and top quarks are taken on-shell\footnote{Even if this choice is not kinematically allowed, it simplifies the extraction of the chromomagnetic contribution.} and the Gordon identity for on-shell fermions is applied to arrive at this result. 
The $(\dots)$ in Eq.~\eqref{eq:diag_tg_2t2phi} represent contributions to vector and axial form factors that are completely removed using on-shell renormalisation of the external top fields.
For the  scheme-dependent value of $K_{tG}^{2t2\phi}$
we find
\begin{equation}
K_{tG}^{2t2\phi}   =
-\frac{g_sm_t}{16\sqrt{2}v\pi^2}\times\begin{cases}
1 & \text{(NDR)}\\
\frac{2}{3}& \text{(BMHV)}\;.
\end{cases}
\label{eq:Ktg_phiF}
\end{equation}
A mapping of $\coeff{tG}{}$ from one scheme to the other in the presence of the operators of Eq.~\eqref{eq:Lag2t2phi} is therefore achieved considering the difference
\begin{equation}
   \Delta K_{tG}^{2t2\phi}=K_{tG}^{2t2\phi, \text{NDR}}-K_{tG}^{2t2\phi, \text{BMHV}}= -\frac{g_sm_t}{48\sqrt{2}v\pi^2}\;,
\end{equation}
similarly as in Eq.~\eqref{eq:CtGMap}.

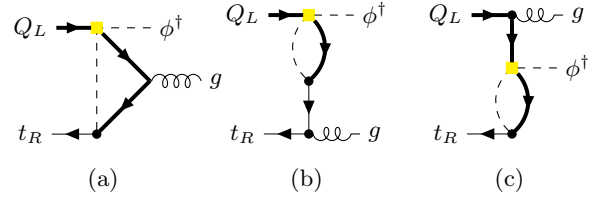
\begin{figure}[t]
\centering
\begin{subfigure}[t]{0.30\linewidth}
\centering
\begin{tikzpicture} 
            \begin{feynman}[small]
                \vertex  (g1)  {$Q_L$};
                \vertex  (gtt1) [square dot, scale=\sizesqdot,right=25pt  of g1,color=yellow] {};
                 \vertex (htt) [dot,scale=0.01,below right = of gtt1]  {};
                \vertex  (gtt2) 
                [dot,scale=\sizedot,below left = of htt] {};
                \vertex  (g2) [left=25pt of gtt2]  {$t_R$};
                \vertex (h) [right = of gtt1] {$\phi^\dagger$};
                \vertex (gg) [right =25pt  of htt] {$g$};

                \diagram* {
                    (g1)  -- [fermion, line width=1.5pt] (gtt1),
                    (gtt2) -- [fermion, line width=0.3pt] (g2),
                    (h)  -- [scalar] (gtt1),
                    (gg)  -- [gluon] (htt),
                    (gtt1) -- [fermion, line width=1.5pt] (htt) -- [fermion, line width=1.5pt]
                    (gtt2)-- [scalar] (gtt1)
                };
            \end{feynman}
        \end{tikzpicture}
        \caption{}
        \end{subfigure}
        \begin{subfigure}[t]{0.30\linewidth}
        \centering
        \begin{tikzpicture} 
            \begin{feynman}[small]
                \vertex  (Q1)  {$Q_L$};
                \vertex  (QQGh1) [square dot, scale=\sizesqdot,right=25pt  of Q1,color=yellow] {};
                \vertex  (QtG1) 
                [dot,scale=\sizedot,below = 25pt of QQGh1] {};
                \vertex  (ttg1) 
                [dot,scale=\sizedot,below = 20pt of QtG1] {};
                \vertex  (t1) [left=25pt of ttg1]  {$t_R$};
                \vertex (h) [right = 25pt of QQGh1] {$\phi^\dagger$};
                \vertex (g1) [right =25pt  of ttg1] {$g$};

                \diagram* {
                    (Q1)  -- [fermion, line width=1.5pt] (QQGh1),
                    (QtG1) -- [fermion, line width=0.3pt] (ttg1)
                     -- [fermion, line width=0.3pt] (t1),
                    (h)  -- [scalar] (QQGh1),
                    (QQGh1) -- [fermion, line width=1.5pt,quarter left] (QtG1)
                     -- [scalar, quarter left] (QQGh1), 
                    (ttg1) -- [gluon] (g1)
                    
                };
            \end{feynman}
        \end{tikzpicture}
        \caption{}
        \end{subfigure}
        \begin{subfigure}[t]{0.30\linewidth}
        \centering
        \begin{tikzpicture} 
            \begin{feynman}[small]
                \vertex  (Q1)  {$Q_L$};
                \vertex  (QQg1) 
                [dot,scale=\sizedot,right=25pt  of Q1] {};
                \vertex  (QQGh1) [square dot, scale=\sizesqdot,below = 20pt of QQg1,color=yellow] {};
                \vertex  (QtG1) 
                [dot,scale=\sizedot,below = 25pt of QQGh1] {};
                \vertex  (t1) [left=25pt of QtG1]  {$t_R$};
                \vertex (h) [right = 25pt of QQGh1] {$\phi^\dagger$};
                \vertex (g1) [right =25pt  of QQg1] {$g$};

                \diagram* {
                    (Q1)  -- [fermion, line width=1.5pt] (QQg1)  -- [fermion, line width=1.5pt] (QQGh1),
                    (QtG1) -- [fermion, line width=0.3pt] (t1),
                    (h)  -- [scalar] (QQGh1),
                    (QQGh1) -- [fermion, line width=1.5pt,quarter left] (QtG1)
                     -- [scalar, quarter left] (QQGh1), 
                    (QQg1) -- [gluon] (g1)
                    
                };
            \end{feynman}
        \end{tikzpicture}
        \caption{}
        \end{subfigure}
        \caption{Contribution to the chromomagnetic operator
with a single insertion of $\mathcal{O}_{ \phi t}$ (yellow square dot) in the unbroken phase.}\label{fig:t2phi2chromo}
    \end{figure}

The same difference is obtained in the unbroken phase, evaluating diagrams of the form of Fig.~\ref{fig:t2phi2chromo} for both operators. 
This provides a solid cross check of the scheme-dependent nature which even holds when the SM gauge bosons are part of the theory, since they cannot contribute to the chromomagnetic operator at one-loop order.

The result of Eq.~\eqref{eq:diag_tg_2t2phi} (and the analogous calculation in the unbroken phase)
illustrates well that we observe a scheme-dependent finite mixing at one-loop between the operators of Eq.~\eqref{eq:Lag2t2phi} and other operators, just like in the case of four-top operators. 
Similarly to Sec.~\ref{sec:matching} a map of finite scheme-dependent shifts in the Wilson coefficients could be verified by an explicit on-shell one-loop matching with an adequate toy model.

Regarding the contribution of those operators to the Higgs-gluon coupling, we refrain from performing the complete calculation as in Sec.~\ref{sec:calculation} in our current work.
Even in the simplified scenario of the gaugeless limit, the contributions of the operators would lead to genuine two-loop Feynman integrals, which is beyond the scope of what we would like to demonstrate here.
With the observed scheme dependence at one-loop, we already expect a $\gamma_5$ scheme dependence for the single pole in $gg\to h$ and for the RGE of $\coeff{\phi G}{}$. As in the case of four-top operators, it should be resolved considering the map of finite shifts in the Wilson coefficients derived at one-loop.
However, it is not guaranteed that the renormalised amplitude of $gg\to h$ would have a scheme-independent form once such shifts are considered.
On the contrary, it may be necessary to identify finite scheme-dependent shifts appearing at the two-loop level.

\section{\label{sec:conclusion}Conclusions}
We have computed the contribution of four-top operators to the Higgs-gluon coupling at two-loop level in the SMEFT. We have discussed in detail, for the first time for this process, the differences between the two schemes for the continuation of $\gamma_5$ to $D$ space-time dimensions considered in this paper, namely NDR and BMHV.
This process is an interesting show-case for the topic of scheme-dependence, because it shows some key features of two-loop computations without adding too many difficulties with
respect to a one-loop computation.

Although the results at two-loop level in the two $\gamma_5$ schemes have a different form, this difference can be accounted for by allowing that the parameters have different values in the two schemes. 
Given this, we determined in Eqs.~(\ref{eq:CtGMap}-\ref{eq:mtMap}) a mapping between the parameters in the two schemes that makes both the anomalous dimension matrix and the finite result scheme-independent. 
This extends the approach presented in Ref.~\cite{Ciuchini:1993ks},
where the scheme-independence of the anomalous dimension matrix only is discussed. To the best of our knowledge, this is the first time that relations such as those in Eqs.~(\ref{eq:CtGMap}-\ref{eq:mtMap}) are considered within the SMEFT. Since the latter is based on the bottom-up approach, Eqs.~(\ref{eq:CtGMap}-\ref{eq:mtMap}) serve an additional purpose compared to Ref.~\cite{Ciuchini:1993ks}, namely they allow to connect the results using different schemes.

We validated the relations between the parameters in the different schemes using some UV models, as detailed in Sec.~\ref{sec:matching}. These simplified UV models support the expectation that the physical result does not depend on the scheme used for $\gamma_5$, if such scheme is used consistently. However, we remark that this holds for a top-down approach, in which the EFT (in this case, the SMEFT) is used as an intermediate step.

In the context of the SMEFT with a new physics scale $\Lambda \sim 1\, \text{TeV}$, the finite terms in the matrix element can be of the same size as the logarithmically enhanced contributions, and thus can be phenomenologically relevant \cite{Alasfar:2022zyr}. For this reason, deriving a connection between the two schemes is very desirable in the perspective of a global fit, where the observables may be computed in different  schemes. To this aim, Eqs.~(\ref{eq:CtGMap}-\ref{eq:mtMap}) represent a first effort in the direction of a comprehensive map between the two schemes. We remark that the continuation scheme for $\gamma_5$ is only one of the calculational choices that could affect the intepretation of SMEFT fits from a bottom-up point of view (see e.g. Refs.~\cite{Corbett:2021cil,Martin:2023fad,Aebischer:2023djt}).

Lastly, we have observed that the interplay of four-top and other SMEFT operators cannot be fully understood in terms of the canonical SMEFT power counting, as in some cases operators that are expected to contribute to different orders based on this counting cannot be treated independently. When the canonical power counting is supplemented by a loop counting like the one discussed in Ref.~\cite{Buchalla:2022vjp}, the observed interplay is more naturally accounted for, under the generic assumption of weakly-coupled and renormalisable UV theories. Furthermore, when the loop counting is considered, the shifts we have presented can be of the same order of magnitude as the Wilson coefficients themselves (see Eq.~\eqref{eq:CtGMap}). As a consequence, experimental constraints on the determination of Wilson coefficients of loop-generated operators (like $\coeff{tG}{}$ in this paper) could be interpreted as suffering from large uncertainties, if scheme-dependent contributions from tree-level-generated chiral operators entering at higher explicit loop orders are omitted (in our case, four-top and $\psi^2 \phi^2 D$ operators).  This points to the necessity of selecting operators contributing to a physical process such that loop counting and canonical-dimension counting are combined, even though it implies assumptions on the UV completion. In any case, a detailed documentation of continuation and renormalisation scheme choices used in EFT calculations and fits of Wilson coefficients is highly recommended.

\section*{Acknowledgments}
We are indebted to Luca Silvestrini, whose comments and suggestions were crucial during the early stages of this project.
We would like to thank various people for discussion: Jorge de Blas, Gerhard Buchalla, Hesham El Faham, Ulrich Haisch, Paride Paradisi, Luca Vecchi and Eleni Vryonidou. We also thank Lina Alasfar for providing assistance in automatising parts of the computation.
The Feynman diagrams shown in this work
were drawn with \texttt{TikZ-Feynman} \cite{Ellis:2016jkw}.
This project has received funding from the European Union’s Horizon Europe research and innovation programme under the Marie Skłodowska-Curie Staff Exchange  grant agreement No 101086085 – ASYMMETRY.  The research of GH and JL was supported by the Deutsche Forschungsgemeinschaft (DFG, German Research Foundation) under grant 396021762 - TRR 257. RG and MV acknowledge support from a departmental research grant under the project “Machine Learning approach to Effective Field Theories in Higgs Physics”. This work is supported in part by the Italian MUR Departments of Excellence grant 2023-2027 "Quantum Frontiers”.
SDN also thanks the Lawrence Berkeley
National Laboratory, Berkeley Center for Theoretical Physics and the Institute for Theoretical Physics at KIT for hospitality.

\appendix
\section{The $h\to b\bar{b}$ rate \label{app:htobb}}
We would like to shortly discuss the computation of the four-quark operators to the $h\to b\bar{b}$ rate both in the NDR and BMHV scheme, which we obtain as a side product of our analysis.
The operators relevant for our discussion are
\begin{equation}
\begin{split}
\mathcal{L}_{\text{b}} &= \frac{\coeff{Qb}{(1)}}{\Lambda^2}  \left(\bar{Q}_L \gamma_\mu Q_L \right) \left(\bar{b}_R \gamma^\mu b_R \right) \\
&+\frac{\coeff{Qb}{(8)}}{\Lambda^2} \left(\bar{Q}_L T^A\gamma_\mu Q_L \right) \left(\bar{b}_R T^A \gamma^\mu b_R \right) \\
&+ \left[ \frac{\coeff{QtQb}{(1)}}{\Lambda^2} \left(\bar{Q}_L   t_R \right) i \tau_2 \left(\bar{Q}_L^{T}  b_R \right)+ \hc \right] \\ 
&+ \left[ \frac{\coeff{QtQb}{(8)}}{\Lambda^2} \left(\bar{Q}_L  T^A  t_R \right) i \tau_2 \left(\bar{Q}_L^{T} T^A  b_R \right)+ \hc \right] \\ 
&+ \left[ \frac{\coeff{b\phi}{}}{\Lambda^2}  (\phi^{\dagger} \phi) \bar{Q}_L \phi b_R+ \hc \right]\,.
\end{split}  
\end{equation}
We consider also scalar operators $\mathcal{O}_{QbQt}^{(1,8)}$ which are neglected in the $gg \to h$ computation since they are suppressed by a factor of $m_b / m_t$.
Including the above operators at NLO, the Higgs decay to bottom quarks is given by Ref.~\cite{Gauld:2015lmb}\footnote{In this reference, the on-shell renormalisation scheme is employed. For this reason, we perform the check with the bare amplitude, Eqs.~(4.13),~(4.14).}

\begin{equation}
\begin{aligned}
\frac{\Gamma_{h\to b\bar{b}}^{\text{NDR}}}{\Gamma_{h\to b\bar{b}}^{\text{SM}}}&=
1 - \frac{m_t}{m_b}  \frac{m_h^2 }{32\pi ^2   \Lambda ^2} \left(7 \coeff{QtQb}{(1)}+\frac{4}{3}  \coeff{QtQb}{(8)} \right)\\ 
&\times \bigg(2 \beta ^3 \log \left(\frac{\beta-1}{\beta + 1}\right)-5 \beta ^2 \\ 
&+\left(1-3 \beta ^2 \right) \log \left(\frac{\tilde{\mu}^2}{m_t^2}\right)+1 \bigg) \\ 
& - \frac{m_h^2}{16 \pi ^2 \Lambda ^2}  \left( \coeff{Qb}{(1)}+ \frac{4}{3}\coeff{Qb}{(8)} \right) \bigg(4 \beta_b^3 \log \left(\frac{\beta_b-1}{\beta_b+1}\right) \\ 
&+7 \beta_b^2+\left(6 \beta_b^2-2\bigg) \log \left(\frac{\tilde{\mu}^2}{m_b^2}\right)-1\right) + \order{\frac{1}{\Lambda^4}} \, ,
\end{aligned}
\end{equation}
and $\beta$ defined in Eq.~\eqref{eq:beta} and $\beta_b$ is obtained from $\beta$ by replacing $m_t$ with $m_b$. The correct branch of the logarithm can be obtained by $m_h^2\to m_h^2+i 0$.
In the BMHV scheme instead the result of the scalar operators does not change with respect to the NDR scheme, but we obtain a different result for the operators $\coeff{Qb}{(1)}$ and $\coeff{Qb}{(8)}$.
We find
\begin{equation}
\frac{\Gamma_{h\to b\bar{b}}^{\text{NDR}}-\Gamma_{h\to b\bar{b}}^{\text{BMHV}}}{\Gamma_{h\to b\bar{b}}^{\text{SM}}}=\frac{\coeff{Qb}{(1)}+ \frac{4}{3} \coeff{Qb}{(8)}}{8\pi^2 \,\Lambda^2}(m_h^2-6 m_b^2) + \order{\frac{1}{\Lambda^4}}\,.
\end{equation}
At tree-level {(TL)} one has $\Gamma_{h\to b\bar{b}}^{\text{X,TL}} \propto (g_{h \bar{b}b}^{\text{X}})^2$, being X=NDR,BMHV, where $g_{h \bar{b}b}$ contains corrections from the operator $\mathcal{O}_{b \phi}$, as can be seen from Eq.~\eqref{eq:ghttmtBF} (replacing $t$ with $b$). 
Keeping into account the different value of such coupling in the two regularisation schemes, namely Eq.~\eqref{eq:ghttMap}, we can write
\begin{align}
\frac{\Gamma_{h\to b\bar{b}}^{\text{NDR,TL}}-\Gamma_{h\to b\bar{b}}^{\text{BMHV,TL}}}{\Gamma_{h\to b\bar{b}}^{\text{SM}}} = & \frac{\coeff{Qb}{(1)}+ \frac{4}{3} \coeff{Qb}{(8)}}{8\pi^2 \, \Lambda^2}(6 m_b^2-m_h^2) \nonumber \\  +& \order{\frac{1}{\Lambda^4}}\,.
\end{align}
If one consistently accounts for the orders in the loop expansion and the $1/\Lambda^2$ expansion, one is then able to obtain a scheme-independent result for this process. 

\section{Renormalisation Group Equations and counterterms\label{app:RGE}}
The anomalous dimension matrix of a theory is strictly connected to the structure of the divergences of the theory itself. In this appendix we analyse in detail this relation, deriving a general formula which can be used to determine the one-loop counterterms associated to SMEFT operators by simply reading the corresponding entry of the renormalisation group equation, given for example in \cite{rge1,rge2,rge3} (or viceversa).

We present here a general argument where a generic SMEFT operator $\mathcal{O}_2$ renormalises a different operator $\mathcal{O}_1$. We fix, coherently with the rest of the paper,
\begin{equation}
\coeff{1}{{\text{MS}}}(\mu) = \coeff{1}{(0)} + \delta \coeff{1}{}(\mu),
\end{equation}
\begin{equation}\label{eq:ct1}
\delta \coeff{1}{}(\mu) = \frac{A}{{\epsilon}}\;  \yuk{}(\mu)^{N_{\yuk{}}} \lambda(\mu) ^{N_\lambda} g(\mu) ^{N_g} \coeff{2}{}(\mu).
\end{equation}
In the previous expression, $\mu$ is the renormalisation scale (on which the ${\text{MS}}$ parameters depend) and $\yuk{},\lambda,g$ denote, respectively, a Yukawa coupling, the Higgs quartic coupling and a gauge coupling and $A$ is a number that does not depend on the renormalisation scale (nor implicitly or explicitly).  

When dimensional regularisation is used, it is customary to rescale the parameters in such a way they maintain their physical dimension: $X \to \mu^{\kappa_X {\epsilon}} X$. A typical example is given by gauge couplings, for which $\kappa_g = 1$ is chosen to keep them dimensionless ($g \to \mu^{{\epsilon }} g$). This operation should be done also for the coefficients of the SMEFT operators, whose mass dimension in $D$ space-time dimensions is different from ${-2}$.\footnote{Within the notation used in this paper, the coefficients are written as $\coeff{i}{}/\Lambda^2$, being $\coeff{i}{}$ a dimensionless quantity.} Remarkably, SMEFT operators may have a different dimension depending on their field content, even if in the limit $D \to 4$ they all have dimension six. Since the product $\coeff{i}{} \mathcal{O}_i$ must have dimension $D$ one has, in principle, 8 different rescaling factors $\kappa_i$, one for each of the operator classes defined in \cite{dim6smeft}. As we will see at the end of this section, keeping this aspect into account is crucial in order to find the correct relation between counterterms and anomalous dimension entries. 

The renormalisation group equation for $\coeff{1}{}$ can be obtained from (dropping the superscript ${\text{MS}}$ for a better readability)
\begin{equation}
\label{eq:RGE1}
0 = \mu \frac{d \coeff{1}{(0)}(\mu)}{d \mu} =\mu \frac{d}{d \mu} \bigg( \mu^{\kappa_1 \epsilon} (\coeff{1}{}(\mu) - \delta \coeff{1}{}(\mu)) \bigg).
\end{equation} 

Since in the end we will take $D \to 4$, we need the first term of the expansion in the $\beta$-function for each of the parameters contained in the counterterm, namely
\begin{equation}
\label{eq:betaeps}
\mu \frac{d X(\mu)}{ d \mu} \equiv \beta_X = - \kappa_X {\epsilon} + \order{1}.
\end{equation}
Performing the algebra in Eq.~\eqref{eq:RGE1} and using Eq.~\eqref{eq:betaeps} we obtain 
\begin{equation} \label{eq:RGE2}
\begin{aligned}
\mu \frac{d \coeff{1}{}(\mu)}{ d \mu } =& A \times (\kappa_1 - \kappa_2 - N_{\yuk{}} - N_g - 2 N_\lambda) \\ 
 & \times \yuk{}(\mu)^{N_{\yuk{}}} \lambda(\mu) ^{N_\lambda} g(\mu) ^{N_g} \coeff{2}{}(\mu) .
\end{aligned}
\end{equation}
If we normalise the anomalous dimension matrix as 
\begin{equation}
\mu \frac{d \coeff{1}{}(\mu)}{ d \mu } = \frac{1}{16 \pi^2} \gamma_{12}(\mu) \coeff{2}{}(\mu),
\end{equation} 
we can write (comparing this expression with Eq.~\eqref{eq:ct1})\footnote{A similar formula taking explicitly into account the rescaling factor to keep the Wilson coefficients with their physical dimension can be found in Ref.~\cite{Adel:1994my} in the context of $b \to s$ transitions.}
\begin{equation}\label{eq:ctfromRGE}
\delta \coeff{1}{} (\mu) = \frac{1}{16 \pi^2 {\epsilon}} \gamma_{12} (\mu)\coeff{2}{}(\mu) \frac{1}{\kappa_1 - \kappa_2 - N_{\yuk{}} - N_g - 2 N_\lambda}.
\end{equation}
A practical example of this formula is Eq.~\eqref{eq:Ctphict}. Four-top operators $\mathcal{O}_{Qt}^{(1,8)}$ renormalise $\mathcal{O}_{t \phi}$ at $\order{\yuk{t } \lambda}$ \cite{rge1} and at $\order{\yuk{t}^3}$ \cite{rge2}. 
This means $(N_{\yuk{}},N_g,N_\lambda)=(1,0,1)$ ($(3,0,0)$) for the former (latter) case. \newline 
In $D=4-2\epsilon$ space-time dimensions one has
\begin{equation}
\dim [\coeff{Qt}{(1,8)}] = 2 \epsilon, \quad \dim [\coeff{t \phi}{}] = 3 \epsilon,
\end{equation}
which implies $\kappa_{Qt} = 2,\, \kappa_{t \phi} = 3$. \newline 
Plugging these numbers in Eq.~\eqref{eq:ctfromRGE} gives Eq.~\eqref{eq:Ctphict} (for both terms of $\order{\yuk{t } \lambda},\,\order{\yuk{t}^3}$).

\section{Additional results} \label{app:results}
We present in this appendix $\mathcal{A}_{g_{h \bar{t}t}+m_t}^{\text{S.I.}}$ introduced in Eq.~\eqref{eq:Amtghtt}

\begin{equation}\label{eq:MSImtghtt}
\begin{aligned}
\mathcal{A}_{g_{h \bar{t}t}+m_t}^{\text{S.I.}}&=-\frac{ g_{h \bar{t}t} g_s^2 m_t }{64 \pi ^4 m_h^4}  L^{\mu_1 \mu_2} \epsilon_{\mu_1}(p_1) \epsilon_{\mu_2} (p_2)  \delta ^{A_1A_2} \\ 
&\times \Bigg[-4 \left(\log \left( \frac{\tilde{\mu}^2}{m_t^2} \right)+2\right) m_h^4 \\ 
&-4 \beta\, m_h^2  \log \left(\frac{\beta-1}{\beta + 1}\right)  \\
&\times \left(2 \left(\log \left( \frac{\tilde{\mu}^2}{m_t^2} \right)-1 \right) m_t^2+m_h^2\right)\\
&+16 \left(2 \log \left( \frac{\tilde{\mu}^2}{m_t^2} \right)+3 \right) m_h^2 m_t^2 \\
&+\log ^2\left(\frac{\beta -1}{\beta +1}\right) \Bigg( \left(\log \left( \frac{\tilde{\mu}^2}{m_t^2} \right)+2 \right) m_h^4 \\
&-4 \left(3 \log \left( \frac{\tilde{\mu}^2}{m_t^2} \right)+5 \right) m_h^2 m_t^2 \\
&+16 \left( 3 \log \left( \frac{\tilde{\mu}^2}{m_t^2} \right)+4 \right) m_t^4 \Bigg) \\
&+ \beta  \log ^3\left(\frac{\beta -1}{\beta + 1}\right) \left(m_h^2-4 m_t^2\right)^2 \Bigg].
\end{aligned}
\end{equation}
$L^{\mu_1 \mu_2}$ has been defined in Eq.~\eqref{eq:Lorentz}, $\beta$ in Eq.~\eqref{eq:beta}, $A_1,A_2$ are the colour indices of the gluons. 
We also report here the result for the SM amplitude for $gg \to h$ at one-loop level:
\begin{equation} \label{eq:MSM}
\begin{aligned}
\mathcal{M}_{\text{SM}} & = \frac{g_{h \bar{t}t} g_s^2}{32 \pi^2 m_t} \tau L^{\mu_1 \mu_2}  \epsilon_{\mu_1}(p_1) \epsilon_{\mu_2} (p_2)  \delta ^{A_1A_2} \\
& \times \left(\beta ^2  \log^2\left(\frac{\beta -1}{\beta +1}\right)-4\right).
\end{aligned}
\end{equation}

\section{Feynman rules} \label{app:FR}
We follow Ref.~\cite{Dedes:2017zog} for what concerns the Feynman rules. For the sake of completeness, we report here the Feynman rules we used in Sec.~\ref{sec:preliminary}:

\begin{align}
        \begin{tikzpicture}[baseline=(4F)]
            \begin{feynman}[small]
                \vertex  (g1)  {$g$};
                 \vertex (gtt1) [square dot, scale=\sizesqdot, right= 30 pt of g1, color=blue] {};
                \vertex  (t1) [above right= 25 pt of gtt1] {$t$};
                \vertex (t2) [below right= 25 pt of gtt1] {$t$};
                \diagram* {
                    (g1)  -- [gluon, momentum = $p$] (gtt1),  
                    (t2) -- [fermion] (gtt1) -- [fermion] (t1),};
            \end{feynman}
        \end{tikzpicture} &= -  \frac{ \coeff{tG}{} }{\Lambda^2}\sqrt{2} v   T^A \sigma^{\mu \nu} p_{\nu},   \label{eq:FR_tg}\\
\begin{tikzpicture}[baseline=(4F)]
            \begin{feynman}[small]
                \vertex  (g1)  {$h$};
                 \vertex (gtt1) [dot, scale=\sizedot, right= 25 pt of g1, color=black] {};
                \vertex  (t1) [above right= 25 pt of gtt1] {$t$};
                \vertex (t2) [below right= 25 pt of gtt1] {$t$};
                \diagram* {
                    (g1)  -- [scalar] (gtt1),  
                    (t2) -- [fermion] (gtt1) -- [fermion] (t1),};
            \end{feynman}
        \end{tikzpicture} &= - i g_{h \bar{t}t},\label{eq:FR_ghtt} \\
\begin{tikzpicture}[baseline=(t1)]
\begin{feynman}[small]
    \vertex  (t1) [] {$t$}; 
    \vertex (4F) [dot, scale=0.01, right = 25 pt of t1,color=black] {};
    \vertex (t2) [right= 25 pt of 4F] {$t$};
    \node[shape=star,star points=4,star point ratio = 15,fill=black, draw,scale = 0.05, rotate=45] at (4F) {};
    \diagram* {
    (t1)  -- [fermion] (4F) -- [fermion] (t2),
     };
\end{feynman}
\end{tikzpicture} &= - i m_t.\label{eq:FR_mt} 
\end{align}
We stress that in Eq.~\eqref{eq:FR_mt} there is not a direct proportionality to the inverse propagator structure $\slashed{p}- \mathbb{1} m_t$, but only to $\mathbb{1} m_t$. For this reason, we added a cross in the fermion line.

Finally, we give the Feynman rule we used for the four-top vertex in Eq.~\eqref{eq:FR_4t}. 
For this rule we explicitly write $s_i,\, c_i$ (spin and color index of the $i$-th quark) to avoid confusion.

In $D=4$ one has 
\begin{equation}
J_{L,\mu} = \gamma_\mu^{(4)} \frac{1-\gamma_5}{2}, \quad J_{R,\mu} = \gamma_\mu^{(4)} \frac{1+\gamma_5}{2}. 
\end{equation}
As detailed in Sec.~\ref{sec:gamma5}, when NDR is used, the continuation $4 \to D$ can be accounted for by replacing $\gamma_\mu ^{(4)} \to \gamma_\mu ^{(D)}$. In BMHV, instead, we follow  ~Refs.~\cite{Ciuchini:1993vr, Belusca-Maito:2023wah, Cornella:2022hkc} for the continuation to $D$ dimensions in order to preserve the chirality of external states.

Following Eq.~\eqref{eq:symmrule}, we have:
\begin{equation}
J_{L,\mu} = \begin{cases}
\gamma^{(D)}_\mu \frac{1-\gamma_5}{2} \qquad \qquad \qquad \quad \quad \text{(NDR)}, \\
\frac{1+\gamma_5}{2} \gamma^{(D)}_\mu \frac{1-\gamma_5}{2} =  \gamma^{(4)}_\mu \frac{1-\gamma_5}{2} \quad \text{(BMHV)},
\end{cases}
\end{equation}
\begin{equation}
J_{R,\mu} = \begin{cases}
\gamma^{(D)}_\mu \frac{1+\gamma_5}{2} \qquad \qquad \qquad \quad \quad \text{(NDR)}, \\
\frac{1-\gamma_5}{2} \gamma^{(D)}_\mu \frac{1+\gamma_5}{2} =  \gamma^{(4)}_\mu \frac{1+\gamma_5}{2} \quad \text{(BMHV)}.
\end{cases}
\end{equation}

\begin{widetext}
 \begin{minipage}{0.2\textwidth}
        \begin{equation*}
\begin{tikzpicture}[baseline=(4F)]
\begin{feynman}[small]
                 \vertex (4t) [square dot, scale=\sizesqdot, color=red] {};
                \vertex  (t1) [above left= 35 pt of 4t] {$t_1$};
                \vertex (t2) [above right= 35 pt of 4t] {$t_2$};
                \vertex  (t3) [below right = 35 pt of 4t] {$t_3$};
                \vertex (t4) [below left= 35 pt of 4t] {$t_4$};

                \diagram* {
                    (t2) -- [fermion] (4t) -- [fermion] (t1),
                    (t4) -- [fermion] (4t) -- [fermion] (t3)
                    };
            \end{feynman}
\end{tikzpicture} =        \end{equation*}
    \end{minipage}%
    \begin{minipage}{0.75\textwidth}
        \begin{equation}
        \begin{aligned}
&+\frac{2 i }{\Lambda^2}\left(\coeff{QQ}{(1)}+\coeff{QQ}{(3)} \right) \times \bigg[ (J_L^\mu)_{s_1 s_2} (J_{L,\mu})_{s_3 s_4} \delta_{c_1 c_2} \delta_{c_3 c_4} - (J_L^\mu)_{s_1 s_4} (J_{L,\mu})_{s_3 s_2} \delta_{c_1 c_4} \delta_{c_3 c_2}   \bigg] \\ 
&+\frac{2 i}{\Lambda^2}\coeff{tt}{} \times \bigg[ (J_R^\mu)_{s_1 s_2} (J_{R,\mu})_{s_3 s_4} \delta_{c_1 c_2} \delta_{c_3 c_4} - (J_R^\mu)_{s_1 s_4} (J_{R,\mu})_{s_3 s_2} \delta_{c_1 c_4} \delta_{c_3 c_2}   \bigg] \\ 
&+ \frac{i}{\Lambda^2} \coeff{Qt}{(1)}\times \bigg[ \bigg( (J_R^\mu)_{s_1 s_2} (J_{L,\mu})_{s_3 s_4} + (J_L^\mu)_{s_1 s_2} (J_{R,\mu})_{s_3 s_4}\bigg) \delta_{c_1 c_2} \delta_{c_3 c_4}\bigg] \\
&- \frac{i}{\Lambda^2} \coeff{Qt}{(1)}\times \bigg[ \bigg( (J_R^\mu)_{s_1 s_4} (J_{L,\mu})_{s_3 s_2} + (J_L^\mu)_{s_1 s_4} (J_{R,\mu})_{s_3 s_2}\bigg) \delta_{c_1 c_4} \delta_{c_3 c_2}\bigg] \\
&+ \frac{i}{\Lambda^2} \coeff{Qt}{(8)}\times \bigg[ \bigg( (J_R^\mu)_{s_1 s_2} (J_{L,\mu})_{s_3 s_4} + (J_L^\mu)_{s_1 s_2} (J_{R,\mu})_{s_3 s_4}\bigg) T^A_{c_1 c_2} T^A_{c_3 c_4}\bigg] \\
&- \frac{i}{\Lambda^2} \coeff{Qt}{(8)}\times \bigg[ \bigg( (J_R^\mu)_{s_1 s_4} (J_{L,\mu})_{s_3 s_2} + (J_L^\mu)_{s_1 s_4} (J_{R,\mu})_{s_3 s_2}\bigg) T^A_{c_1 c_4} T^A_{c_3 c_2}\bigg] 
\end{aligned}
\label{eq:FR_4t}
\end{equation}
    \end{minipage}
\end{widetext}

\vfill
\bibliography{bibliography}

\end{document}